%% file: survey.tex
\documentclass[3p,times,twocolumn]{elsarticle}
\usepackage{ctable}
\usepackage{acronym}
\usepackage{multirow}
\usepackage{xspace}
\usepackage{booktabs} 
\usepackage{longtable}
\usepackage{lscape}
\usepackage{balance}

\usepackage[ruled]{algorithm2e} 
\SetAlFnt{\small}
\SetAlCapFnt{\small}
\SetAlCapNameFnt{\small}
\SetAlCapHSkip{0pt}
\IncMargin{-\parindent}

\usepackage[hyphens]{url}

\usepackage{hyperref}
\hypersetup{
    pdfnewwindow=true,      
    colorlinks=true,       
    urlcolor=blue           
}

\graphicspath{{./}}

\acrodef{QoS}{Quality of Service}
\acrodef{VM}{Virtual Machine}
\acrodef{UDF}{User Defined Function}
\acrodef{DAG}{Directed Acyclic Graph}
\acrodef{CEP}{Complex Event Processing}
\acrodef{S4}{Simple Scalable Streaming System}
\acrodef{OWL}{Web Ontology Language}
\acrodef{SPC}{Stream Processing Core}
\acrodef{CEP}{Complex Event Processing}
\acrodef{DSMS}{Data Stream Management System}
\acrodef{SPE}{Stream Processing Engine}
\acrodef{DBMS}{DataBase Management System} 
\acrodef{JSON}{JavaScript Object Notation}
\acrodef{CQL}{Continuous Query Language}
\acrodef{PE}{Processing Element}
\acrodef{IoT}{Internet of Things}
\acrodef{RDD}{Resilient Distributed Dataset}
\acrodef{MAPE}{Monitoring, Analysis, Planning and Execution}
\acrodef{ETL}{Extract, Transform, and Load}
\acrodef{SDN}{Software Defined Network}
\acrodef{NFV}{Network Functions Virtualization}
\acrodef{DHT}{Distributed Hash Table}

\acrodef{PaaS}{Platform as a Service}
\acrodef{IaaS}{Infrastructure as a Service}
\acrodef{SLO}{Service Level Objective}
\acrodef{ETP}{Effective Throughput Percentage}
\acrodef{DSPS}{Data Stream-processing System}
\acrodef{Stela}{STream processing ELAsticity}
\acrodef{SBON}{Stream-Based Overlay Network}

\acrodef{DSP}{Data Stream Processing}
\acrodef{ODP}{Optimal Data Stream Processing Placement}
\acrodef{CSA}{Connected Streaming Analytics}

\acrodef{WSN}{Wireless Sensor Network}
\acrodef{WAN}{Wide Area Network}
\acrodef{SLA}{Service Level Agreement}
\acrodef{VISP}{VIenna ecosystem for elastic Stream Processing}
\acrodef{ASA}{Azure Stream Analytics}

\acrodef{SPL}{Stream Processing Language}
\acrodef{MQTT}{Message Queue Telemetry Transport}

\acrodef{AWS}{Amazon Web Services}
\acrodef{EC2}{Elastic Compute Cloud}
\acrodef{ES}{Elasticsearch Service}
\acrodef{S3}{Simple Storage Service}
\acrodef{AmazonECS}[ECS]{EC2 Container Service}

\newcommand{\tuple}[1]{\ensuremath{\left \langle #1 \right \rangle }}

\newcommand{\etal}{\textit{et al.}\xspace}
\newcommand{\ie}{\textit{i.e.}\xspace}
\newcommand{\eg}{\textit{e.g.}\xspace}

\begin{document}

\begin{frontmatter}

\title{Distributed Data Stream Processing and Edge Computing: \\A Survey on Resource Elasticity and Future Directions}

\author[inria]{Marcos Dias de Assun\c{c}\~{a}o\corref{cor1}}
\author[inria]{Alexandre da Silva Veith}
\author[unimelb]{Rajkumar Buyya}

\address[inria]{Inria, LIP, ENS Lyon, France}
\address[unimelb]{The University of Melbourne, Australia}
\cortext[cor1]{Corresponding author: assuncao@acm.org}

\begin{abstract}
Under several emerging application scenarios, such as in smart cities, operational monitoring of large infrastructure, wearable assistance, and Internet of Things, continuous data streams must be processed under very short delays. Several solutions, including multiple software engines, have been developed for processing unbounded data streams in a scalable and efficient manner. More recently, architecture has been proposed to use edge computing for data stream processing. This paper surveys state of the art on stream processing engines and mechanisms for exploiting resource elasticity features of cloud computing in stream processing. Resource elasticity allows for an application or service to scale out/in according to fluctuating demands. Although such features have been extensively investigated for enterprise applications, stream processing poses challenges on achieving elastic systems that can make efficient resource management decisions based on current load. Elasticity becomes even more challenging in highly distributed environments comprising edge and cloud computing resources. This work examines some of these challenges and discusses solutions proposed in the literature to address them.       
\end{abstract}


\end{frontmatter}

\input{introduction}
\input{background}
\input{spes}
\input{cloud_systems}
\input{elasticity}
\input{architectures}

\input{gap_analysis}

\section{Summary and Conclusions}
\label{sec:conclusions}

This paper discussed solutions for stream processing and techniques to manage resource elasticity. It first presented how stream processing fits in the overall data processing framework often employed by large organisations. Then it presented a historical perspective on stream processing engines, classifying them into three generations. After that, we elaborated on third-generation solutions and discussed existing work that aims to manage resource elasticity for stream processing engines. In addition to discussing the management of resource elasticity, we highlighted the challenges inherent to adapting stream processing applications dynamically in order to use additional resources made available during scale out operations, or release unused capacity when scaling in. The work then discussed emerging distributed architecture for stream processing and future directions on the topic. We advocate the need for high-level programming abstractions that enable developers to program and deploy stream processing applications on these emerging and highly distributed architecture more easily, while taking advantage of resource elasticity and fault tolerance.  

\section*{Acknowledgements}

We thank Rodrigo Calheiros (Western Sydney University), Srikumar Venugopal (IBM Research Ireland), Xunyun Liu (The University of Melbourne), and Piotr Borylo (AGH University) for their comments on a preliminary version of this work. This work has been carried out in the scope of a joint project between the French National Center for Scientific Research (CNRS) and the University of Melbourne.

\balance
\bibliographystyle{elsarticle-num}
\bibliography{references}
\end{document}

%% file: introduction.tex
\section{Introduction}

The increasing availability of sensors, mobile phones, and other devices has led to an explosion in the volume, variety and velocity of data generated and that requires analysis of some type. As society becomes more interconnected, organisations are producing vast amounts of data as result of instrumented business processes, monitoring of user activity \cite{CiscoMSE:2012,attentionshoppers:2013}, wearable assistance \cite{HaWearable:2014}, website tracking, sensors, finance, accounting, large-scale scientific experiments, among other reasons. This data deluge is often termed as \textit{big data} due to the challenges it poses to existing infrastructure regarding, for instance, data transfer, storage, and processing \cite{AssuncaoJPDC:2015}.  

A large part of this big data is most valuable when it is analysed quickly, as it is generated. Under several emerging application scenarios, such as in smart cities, operational monitoring of large infrastructure, and \ac{IoT} \cite{AtzoriIoTSurvey:2010}, continuous data streams must be processed under very short delays. In several domains, there is a need for processing data streams to detect patterns, identify failures \cite{RettigOnline:2015}, and gain insights.

Several stream processing frameworks and tools have been proposed for carrying out analytical tasks in a scalable and efficient manner. Many tools employ a dataflow approach where incoming data results in data streams that are redirected through a directed graph of operators placed on distributed hosts that execute algebra-like operations or user-defined functions. Some frameworks, on the other hand, discretise incoming data streams by temporarily storing arriving data during small time windows and then performing micro-batch processing whereby triggering distributed computations on the previously stored data. The second approach aims at improving the scalability and fault-tolerance of distributed stream processing tools by handling straggler tasks and faults more efficiently.   

Also to improve scalability, many stream processing frameworks have been deployed on clouds \cite{ArmbrustCloud:2009}, aiming to benefit from characteristics such as resource elasticity. Elasticity, when properly exploited, refers to the ability of a cloud to allow a service to allocate additional resources or release idle capacity on demand to match the application workload. Although efforts have been made towards making stream-processing more elastic, many issues remain unaddressed. There are challenges regarding the placement of stream processing tasks on available resources, identification of bottlenecks, and application adaptation. These challenges are exacerbated when services are part of a larger infrastructure that comprises multiple execution models (\eg lambda architecture, workflows or resource-management bindings for high-level programming abstractions \cite{BoykinSummingbird:2014,GoogleDataFlow}) or hybrid environments comprising both cloud and edge computing resources \cite{ChaoETSI:2015,HuEdge:2016}.

More recently, software frameworks \cite{ApacheEdgent,PisaniSBAC:2017} and architectures have been proposed for carrying out data stream processing using constrained resources located at the edge of the Internet. This scenario introduces additional challenges regarding application scheduling, resource elasticity, and programming models. This article surveys stream-processing solutions and approaches for deploying data stream processing on cloud computing and edge environments. By so doing, it makes the following contributions:

\begin{itemize}
\item It reviews multiple generations of data stream processing frameworks, describing their architectural and execution models.   
\item It analyses and classifies existing work on exploiting elasticity to adapt resource allocation to match the demands of stream processing services. Previous work has surveyed stream processing solutions without a focus on how resource elasticity is addressed \cite{ZhaoSurveyStreams:2017}. The present work provides a more in-depth analysis of existing solutions and discusses how they attempt to achieve resource elasticity.
\item It discusses ongoing efforts on resource elasticity for data stream processing and their deployment on edge computing environments, and outlines future directions on the topic. 
\end{itemize} 

The rest of this paper is organised as follows.
Section \ref{sec:background} provides background information on big-data ecosystems and architecture for online data processing. Section \ref{sec:spes_tools} describes existing engines and other software solutions for data stream processing whereas Section \ref{sec:cloud_systems} discusses managed cloud solutions for stream processing. In Section \ref{sec:elasticity} we elaborate on how existing work tries to tackle aspects of resource elasticity for data stream processing. Section \ref{sec:architectures} discusses solutions that aim to leverage multiple types of infrastructure (\eg cloud and edge computing) to improve the performance of stream processing applications. Section \ref{sec:future_directions} presents future directions on the topic and finally, Section \ref{sec:conclusions} concludes the paper.

%% file: background.tex
\section{Background and Architecture}
\label{sec:background}

This section describes background on stream-processing systems for big-data. It first discusses how layered real-time architecture is often organised and then presents a historical summary of how such systems have evolved over time.  

\subsection{Online Data Processing Architecture}

\begin{figure*}[ht]
\centering 
\includegraphics[width=1.\linewidth]{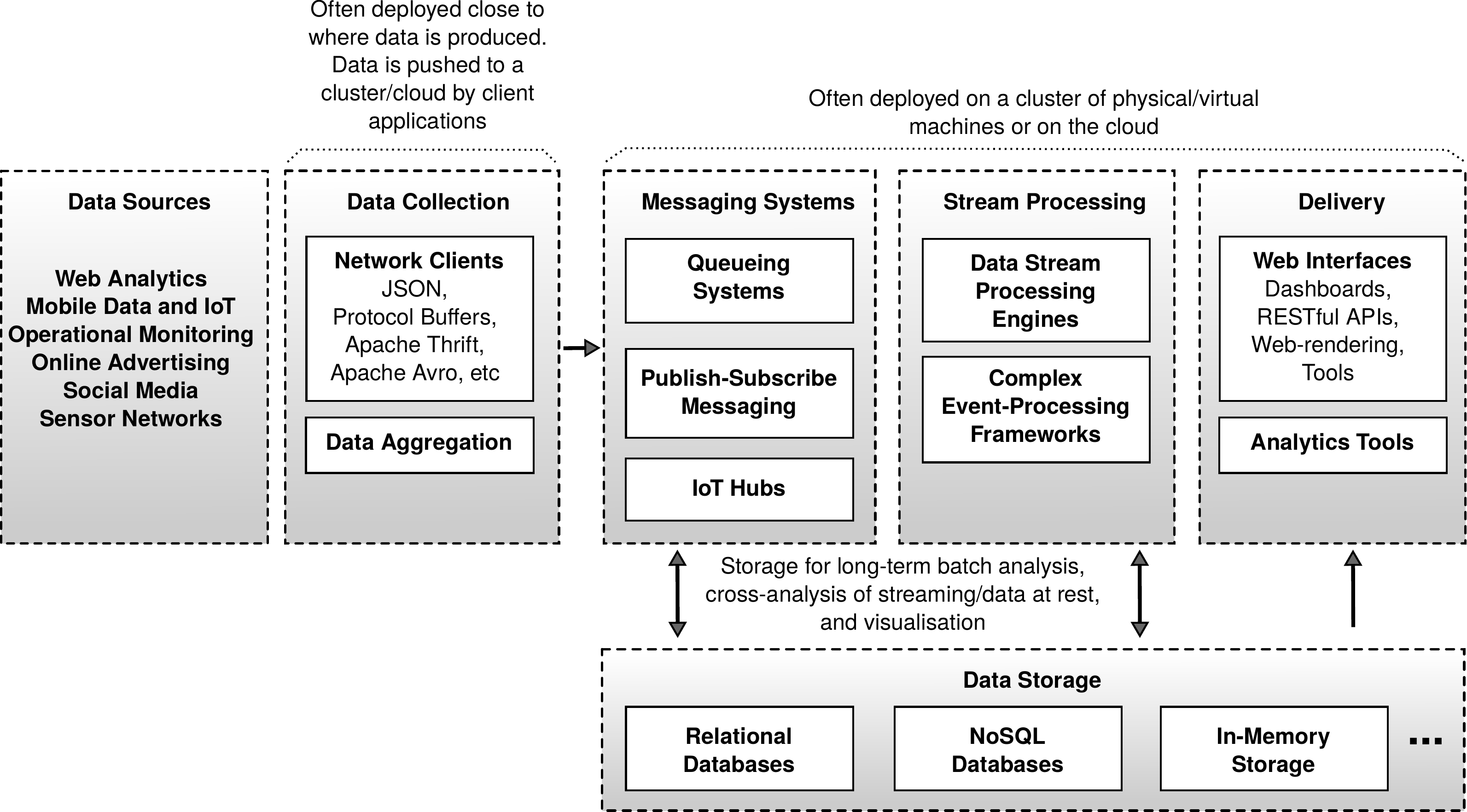} 
\caption{Overview of an online data-processing architecture.}
\label{fig:streaming_architecture}
\end{figure*}

Architecture for online\footnote{Similar to Boykin \etal, hereafter use the term \textit{online} to mean that \textit{``data are processed as they are being generated''}.} data analysis is generally multi-tiered systems that comprise many loosely coupled components \cite{EllisRealTimeStreaming:2014,AllenStormApplied:2015,LiuStreamIoT:2016}. While the reasons for structuring architecture in this way may vary, the main goals include improving maintainability, scalability, and availability. 
Figure~\ref{fig:streaming_architecture} provides an overview of components often found in a stream-processing architecture. Although an actual system might not have all these components, the goal here is to describe how a stream processing architecture may look like and position the stream processing solutions discussed later.

The \textit{Data Sources} (Figure~\ref{fig:streaming_architecture}) that require timely processing and analysis include Web analytics, infrastructure operational monitoring, online advertising, social media, and \ac{IoT}. Most \textit{Data Collection} is performed by tools that run close to where the data and that communicate the data via TCP/IP connections, UDP, or long-range communication \cite{CentenaroIoT:2016}. Solutions such as \ac{JSON} are used as a data-interchange format. For more structured data, wire protocols such as Apache Thrift \cite{ApacheThrift} and Protocol Buffers \cite{ProtoBuff}, can be employed. Other messaging protocols have been proposed for \ac{IoT}, some of which are based on HTTP \cite{AtzoriIoTSurvey:2010}. Most data collection activities are executed at the edges of a network, and some level of data aggregation is often performed via, for instance \ac{MQTT}, before data is passed through to be processed and analysed.

An online data-processing architecture can comprise multiple tiers of collection and processing, with the connection between these tiers made on an ad-hoc basis. To allow for more modular systems, and to enable each tier to grow at different paces and hence accommodate changes, the connection is at times made by message brokers and queuing systems such as Apache ActiveMQ \cite{ApacheActiveMQ}, RabbitMQ \cite{RabiitMQ} and Kestrel \cite{Kestrel}, publish-subscribe based solutions including Apache Kafka \cite{ApacheKafka} and DistributedLog \cite{DistributedLog}, or managed services such as Amazon Kinesis Firehose~\cite{AmazonFirehose} and Azure IoT Hubs~\cite{AzureIoTHub}. These systems are termed here as ``Messaging Systems'' and they enable, for instance, the processing tier to expand to multiple data centres and collection to be changed without impacting processing.

Over the years several models and frameworks have been created for processing large volumes of data, among which MapReduce is one of the most popular \cite{JeffreyMapReduce:2008}. Although most frameworks process data in a batch manner, numerous attempts have been made to adapt them to handle more interactive and dynamic workloads \cite{BorthakurFacebookHadoop:2011,ChenEnergyHadoop:2012}. Such solutions handle many of today's use cases, but there is an increasing need for processing collected data always at higher rates and providing services with short response time. \textit{Data Stream Processing} systems are commonly designed to handle and perform one-pass processing of unbounded streams of data. This tier, the main focus of this paper, includes solutions that are commonly referred to as stream management systems and complex-event processing systems. The next sections review data streams and provide a historic overview of how this core component of the data processing pipeline has evolved over time. 

Moreover, a data processing architecture often stores data for further processing, or as support to presente results to analysts or deliver them to other analytics tools. The range of \textit{Data Storage} solutions used to support a real-time architecture are numerous, ranging from relational databases, to key-value stores, in-memory databases, and NoSQL databases \cite{HanNOSQLSurvey:2011}. The results of data processing are delivered (\ie Delivery tier) to be used by analysts or machine learning and data mining tools. Means to interface with such tools or to present results to be visualised by analysts include RESTful or other Web-based APIs, Web interfaces and other rendering solutions. There are also many data storage solutions provided by cloud providers such as Amazon, Azure, Google, and others.

\subsection{Data Streams and Models}

The definition of a data stream can vary across domains, but in general, it is commonly regarded as input data that arrives at a high rate, often being considered as big data, hence stressing communication and computing infrastructure. The type of data in a stream may vary according to the application scenario, including discrete signals, event logs, monitoring information, time series data, video, among others. Moreover, it is also important to distinguish between streaming data when it arrives at the processing system via, for instance, a log or queueing system, and intermediate streams of \textit{tuples} resulting from the processing by system elements. When discussing solutions, this work focuses on the resource management and elasticity aspects concerning the intermediate streams of tuples created or/and processed by elements of a stream processing system.     

Multiple attempts have been made towards classifying stream types. Muthukrishnan \cite{MuthukrishnanStreams:2005} classifies data streams under several models based on how their input data describes the underlying signal they represent. The identified models include \textit{time series}, \textit{cash register}, and \textit{turnstile}. Many of the application domains envisioned when these models were identified concern operational monitoring and financial markets. More recent streams of data generated by applications such as social networks can be semi-structured or unstructured, thus carrying information about multiple signals. In this work, an input data stream is an online and unbounded sequence of data elements \cite{Babcock:2002,GolabIssuesStream:2003}. The elements can be homogeneous, hence structured, or heterogeneous, thus semi-structured or unstructured. More formally, an input stream is a sequence of data elements $e_1, e_2,\dots$ that arrive one at a time, where each element $e_{i}$ can be viewed as $e_i = (t_i,D_i)$ where $t_i$ is the time stamp associated with the element, and $D_i = \tuple{d_1,d_2,\dots}$ is the element payload, here represented as a \textit{tuple} of data items.

As mentioned earlier, many stream processing frameworks use a data flow abstraction by structuring an application as a graph, generally a \ac{DAG}, of \textit{operators}. These operators perform functions such as counting, filtering, projection, and aggregation, where the processing of an input data stream by an element can result in the creation of subsequent streams that may differ from the original stream in terms of data structure and rate.

\begin{figure}[ht]
\centering 
\includegraphics[width=.8\linewidth]{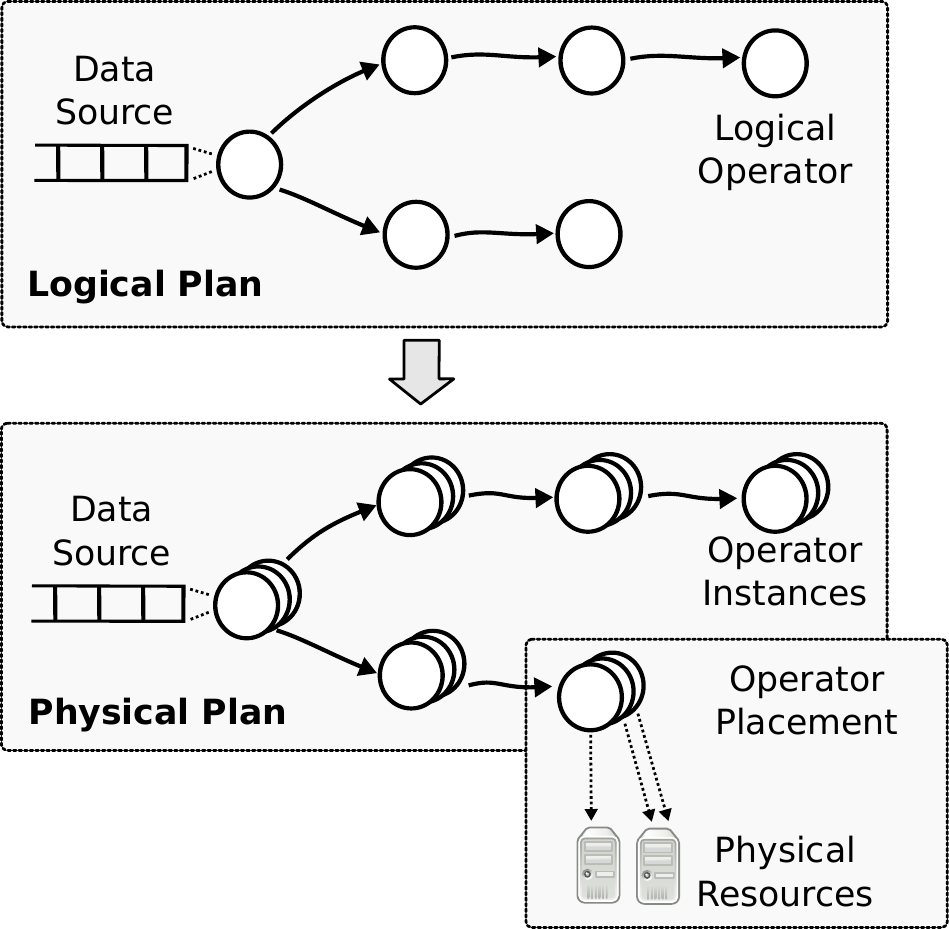} 
\caption{Logical and physical operator plans.}
\label{fig:logical_plan}
\end{figure}

Frameworks that structure data stream processing applications as data flow graph generally employ a logical abstraction for specifying operators and how data flows between them; this abstraction is termed here as \textit{logical plan} \cite{KulkarniHeron:2015} (see Figure~\ref{fig:logical_plan}). As explained in detail later, a developer can provide parallelisation hints or specify how many instances of each operator should be created when building the \textit{physical plan} that is used by a scheduler or another component responsible for placing the operator instances on available cluster resources. As depicted in the figure, physical instances of a same logical operator may be placed onto different physical or virtual resources.  

\begin{figure}[ht]
\centering 
\includegraphics[width=1.\linewidth]{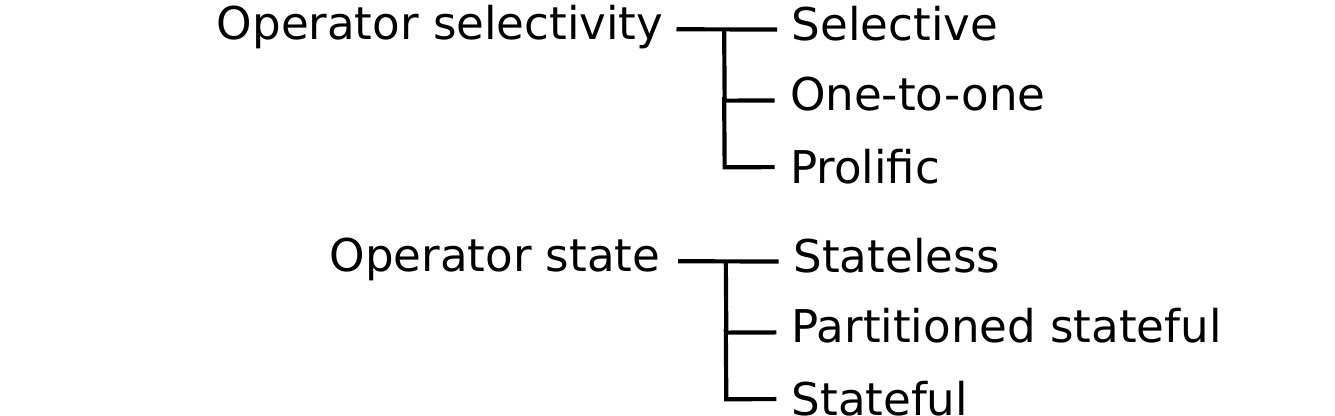} 
\caption{Types of operator selectivity and state.}
\label{fig:op_selectivity}
\end{figure}

With respect to the selectivity of an operator (\ie the number of items it produces per number of items consumed) it is generally classified \cite{GedikPipelinedFission:2016} (Figure~\ref{fig:op_selectivity}) as \textit{selective}, where it produces less than one; \textit{one-to-one}, where the number of items is equal to one; or \textit{prolific}, in which it produces more than one. Regarding state, an operator can be \textit{stateless}, in which case it does not maintain any state between executions; \textit{partitioned stateful} where a given data structure maintains state for each downstream based on a partitioning key, and \textit{stateful} where no particular structure is required.

\begin{figure}[ht]
\centering 
\includegraphics[width=1.\linewidth]{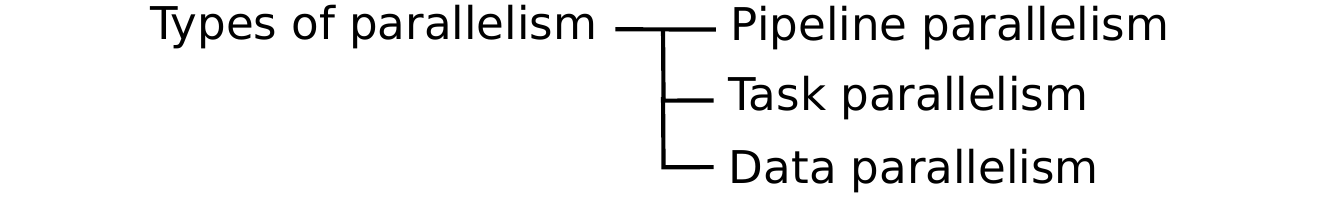} 
\caption{Some types of parallelism enabled by data-flow based stream processing.}
\label{fig:op_parallelism}
\end{figure}

Organising a data stream processing application as a graph of operators allows for exploring certain levels of parallelism (Figure~\ref{fig:op_parallelism}) \cite{TangAutoPipelining:2013}. For example, \textit{pipeline parallelism} enables an operator to process a tuple while an upstream operator can handle the next tuple concurrently. Graphs can contain segments that execute the same set of tuples in parallel, hence exploiting \textit{task parallelism}. Several techniques also aim to use \textit{data parallelism}, which often requires changes in the graph to replicate operators and adjust the data streams between them. For example, parallelising regions of a chain graph  \cite{GedikPipelinedFission:2016} may consist of creating multiple pipelines preceded by an operator that partitions the incoming tuples across the downstream pipelines -- often called a \textit{splitter} -- and followed by an operator that merges the tuples processed along the pipelines -- termed as an \textit{mergers}. Although parallelising regions can increase throughput, they may require mechanisms to guarantee time semantics, which can make splitters and mergers block for some time to guarantee, for instance, time order of events.

\subsection{Distributed Data Stream Processing}

Several systems have been developed to process dynamic or streaming data \cite{SattlerElastic:2013,LiuStreamIoT:2016}, hereafter termed as \acp{SPE}. One of the categories under which such systems fall is often called \ac{DSMS}, analogous to \acp{DBMS} which are responsible for managing disk-resident data usually providing users with means to perform relational operations among table elements. \acp{DSMS} include operators that perform standard functions, joins, aggregations, filtering, and advanced analyses. Early \acp{DSMS} provided SQL-like declarative languages for specifying long-running queries over unbounded streams of data. \ac{CEP} systems \cite{WuCEPLanguage:2006}, a second category, supports the detection of relationships among events, for example, temporal relations that can be specified by correlation rules, such a sequence of specific events over a given time interval. \ac{CEP} systems also provide declarative interfaces using event languages like SASE \cite{GyllstromSASE:2007} or following data-flow specifications.

The first generation of \acp{SPE} provided extensions to the traditional \ac{DBMS} model by enabling long-running queries over dynamic data, and by offering declarative interfaces and SQL-like languages that allowed a user to specify algebra-like operations. Most engines were restricted to a single machine and were not executed in a distributed fashion. The second generation of engines enabled distributed processing by decoupling processing entities that communicate with one another using message-passing processes. This enhanced model could take advantage of distributed hosts, but introduced challenges about load balancing and resource management. Despite the improvements in distributed execution, most engines of these two generations fall into the category of \acp{DSMS}, where queries are organised as operator graphs. IBM proposed System S, an engine based on data-flow graphs where users could develop operators of their own. The goal was to improve scalability and efficiency in stream processing, a problem inherent to most \acp{DSMS}. Achieving horizontal scalability while providing declarative interfaces still remained a challenge not addressed by most engines.

More recently, several \acp{SPE} were developed to perform distributed stream processing while aiming to achieve scalable and fault-tolerant execution on cluster environments. Many of these engines do not provide declarative interfaces, requiring a developer to program applications rather than write queries. Most engines follow a one-pass processing model where the application is designed as a data-flow graph. Data items of an input stream, when received, are forwarded throw a graph of processing elements, which can, in turn, create new streams that are redirected to other elements. These engines allow for the specification of \acp{UDF} to be performed by the processing elements when an application is deployed. Another model that has gained popularity consists in discretising incoming data streams and launching periodical micro-batch executions. Under this model, data received from an input stream is stored during a time window, and towards the end of the window, the engine triggers distributed batch processing. Some systems trigger recurring queries upon bulk appends to data streams \cite{HeComet:2010}. This model aims to improve scalability and throughput for applications that do not have stringent requirements regarding processing delays.

\begin{figure}[ht]
\centering 
\includegraphics[width=1.\linewidth]{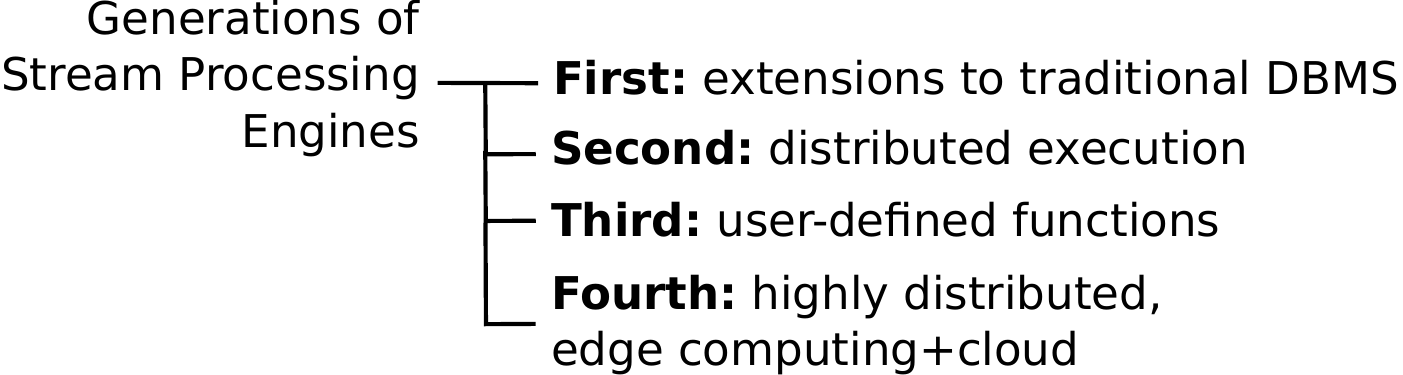} 
\caption{Generations of Stream Processing Engines.}
\label{fig:spe_generations}
\end{figure}

We are currently witnessing the emergence of a fourth generation of data stream processing frameworks, where certain processing elements are placed on the edges of the network. Architectural models \cite{Sajjad:2016}, \acp{SPE} \cite{ChanQuarksIBM:2016,PisaniSBAC:2017}, and engines for certain application scenarios such as \ac{IoT} are emerging. Architecture that mixes elements deployed on edge computing resources and the cloud is provided in the literature \cite{ChanQuarksIBM:2016,HirzelSPL:2017,Sajjad:2016}. 

The generations of \acp{SPE} are summarised in Figure~\ref{fig:spe_generations}. Although we discuss previous generations of \ac{DSMS} and \ac{CEP} solutions, this work focuses on state of the art frameworks and technology for stream processing and solutions for exploiting resource elasticity for stream processing engines that accept \acp{UDF}. We focus on the third generation of stream processing frameworks while discussing some of the challenges inherent to the fourth. 


\subsection{Resource Elasticity}

Cloud computing is a model under which organisations of all sizes can lease IT resources and services on-demand and pay as they go \cite{ArmbrustCloud:2009}. Resources allocated to customers are often \acp{VM} or containers that share the underlying physical infrastructure, which allows for workload consolidation that can hence lead to better system utilisation and energy efficiency. Another important feature of clouds is \textit{resource elasticity}, which enables organisations to change infrastructure capacity dynamically with the support of \textit{auto-scaling operations}. This capability is essential in several settings as it helps service providers: to minimise the number of allocated resources and to deliver adequate \ac{QoS} levels, usually synonymous with low response times.

In addition to deciding when to modify the system capacity, auto-scaling algorithms must identify adequate step-sizes (\ie the number of resources by which the cloud should shrink and expand) during scale out/in operations in order to prevent resource wastage and unacceptable \ac{QoS} \cite{NettoScaling:2014}. An elastic system requires not only mechanisms that adjust service execution to current resource capacity -- \eg present horizontal scalability -- but also an \textit{auto-scaling policy} that defines when and by how much resource capacity is added or removed.

Auto-scaling policies were proposed for several types of enterprise applications and certain big-data workloads, mostly those that process data in batches. Although resource elasticity for stream processing applications has been investigated in previous work, several challenges are not yet fully addressed \cite{SattlerElastic:2013}. As highlighted by Tolosana-Calasanz \etal \cite{TolosanaMultiTenancyStreas:2016}, mechanisms for scaling resources in cloud infrastructure can still incur severe delays.  For stream processing engines that organise applications as operator graphs, an elastic operation that adds more nodes at runtime may require re-routing the data and migrating stream processing operators. Moreover, as stream processing applications run for long periods of time and cannot be restarted without losing data, resource allocation must be performed much more carefully.

When considering solutions for managing elasticity of data streaming, this work discusses the techniques and metrics employed for monitoring the performance of data stream processing systems and the actions carried out during auto-scaling operations. The actions performed during auto-scaling operations include, for instance adding/removing computing resources and adjusting the stream processing application by changing the level of parallelism of certain processing operators, adjusting the processing graph, merging or splitting operators, among other things.

%% file: spes.tex
\acrodef{ELF}{Efficient, Lightweight, Flexible}

\section{Stream Processing Engines and Tools}
\label{sec:spes_tools}

While the first generation of \acp{SPE} were analogous to \acp{DBMS}, developed to perform long running queries over dynamic data and consisted essentially of centralised solutions, the second generation introduced distributed processing and revealed challenges on load balancing and resource management. The third generation of solutions resulted in more general application frameworks that enable the specification and execution of \acp{UDF}. This section presents a historical overview of data stream processing solutions and then discusses third-generation solutions.

\subsection{Early Stream Processing Solutions}

The first-generation of stream processing systems dates back to 2000s and were essentially extensions of \acp{DBMS} for performing continuous queries that, compared to today's scenarios, did not process large amounts of data. In most systems, an application or query is a \ac{DAG} whose vertices are operators that execute functions that transform one or multiple data streams and edges that define how data elements flow from one operator to another. The execution of a function by an operator over an incoming data stream can result in one or multiple output streams. This section provides a select list of these systems and describes their properties.

NiagaraCQ \cite{ChenNiagaraCQ:2000} was conceived to perform two categories of queries over XML datasets, namely queries that are executed as new data becomes available and continuous queries that are triggered periodically. STREAM \cite{ArasuSTREAM:2004} provides a \ac{CQL} for specifying queries executed over incoming streams of structured data records. STREAM compiles \acp{CQL} queries into query plans, which comprise operators that process tuples, queues that buffer tuples, and synopses that store operator state. A query plan is an operator tree or a \ac{DAG}, where vertices are operators, and edges represent their composition and define how the data flows between operators. When executing a query plan, the scheduler selects plan operators and assigns them to available resources. Operator scheduling presents several challenges as it needs to respect constraints concerning query response time and memory utilisation. STREAM uses a chain scheduling technique that aims to minimise memory usage and adapt its execution to variations in data arrival rate \cite{BabcockChain:2003}.

Aurora \cite{AbadiAurora:2003} was designed for managing data streams generated by monitoring applications. Similar to STREAM, it enables continuous queries that are viewed as \acp{DAG} whose vertices are operators, and edges that define the tuple flow between operators. Aurora schedules operators using a technique termed as train scheduling that explores non-linearities when processing tuples by essentially storing tuples at the input of so-called boxes, thus forming a train, and processing them in batches. It pushes tuple trains through multiple boxes hence reducing I/O operations. 

As a second-generation of stream processing systems, Medusa \cite{BalazinskaMedusa:2005} uses Aurora as its query processing engine and arranges its queries to be distributed across nodes, routeing tuples and results as needed. By enabling distributed processing and task migration across participating nodes, Medusa introduced several challenges in load balancing, distribute load shedding \cite{TatbulFIT:2007}, and resource management. For instance, the algorithm for selecting tasks to offload must consider the data flow among operators. Medusa offers techniques for balancing the load among nodes, including a contract-based scheme that provides an economy-inspired mechanism for overloaded nodes to shed tasks to other nodes. Borealis \cite{AbadiBorealis:2005} further extends the query functionalities of Aurora and the processing capabilities of Medusa \cite{BalazinskaMedusa:2005} by dynamically revising query results, enabling query modification, and distributing the processing of operators across multiple sites. Medusa and Borealis have been key to distributed stream processing, even though their operators did not allow for the execution of user-defined functions, a key feature of current stream processing solutions.

\subsection{Current Stream Processing Solutions}

\begin{figure}[ht]
\centering 
\includegraphics[width=.85\linewidth]{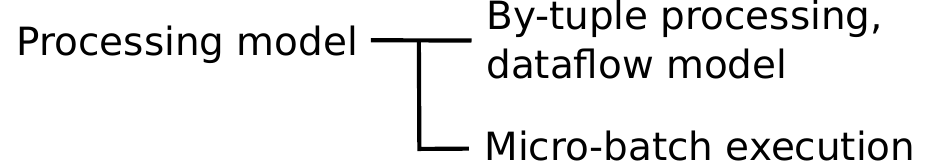} 
\caption{Streaming processing approaches.}
\label{fig:processing_model}
\end{figure}

Current systems enable the processing of unbounded data streams across multiple hosts and the execution of \acp{UDF}. Numerous frameworks have been proposed for distributed processing following essentially two models (Figure \ref{fig:processing_model}): 
\begin{itemize}
\item the \textit{operator-graph} model described earlier, where a processing system is continuously ingesting data that is processed at a by-tuple level by a \ac{DAG} of operators; and 
\item a \textit{micro-batch} in which incoming data is grouped during short intervals, thus triggering a batch processing towards the end of a time window. The rest of this section provides a description of select systems that fall into these two categories. 
\end{itemize}

\subsubsection{Apache Storm}
An application in Storm, also called a \textit{Topology}, is a computation graph that defines the processing elements (\textit{i.e. Spouts} and \textit{Bolts}) and how the data (\textit{\ie tuples}) flows between them. A topology runs indefinitely, or until a user stops it. Similarly to other application models, a topology receives an influx of data and divides it into chunks that are processed by tasks assigned to cluster nodes. The data that nodes send to one another is in the form of sequences of \textit{Tuples}, which are ordered lists of values.

\begin{figure}[ht]
\centering 
\includegraphics[width=1.\linewidth]{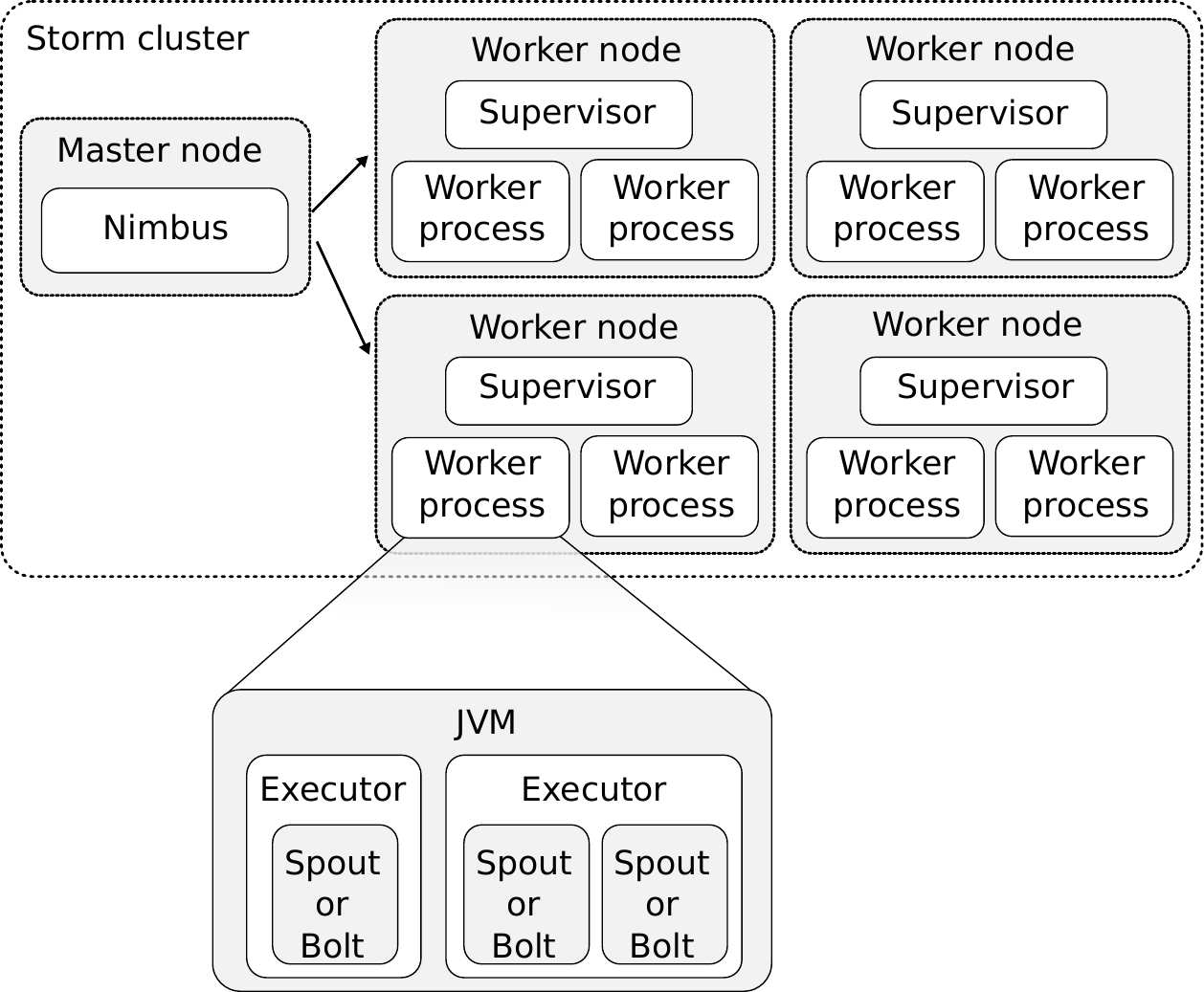} 
\caption{Main components of a Storm cluster \cite{AllenStormApplied:2015}.}
\label{fig:storm_architecture}
\end{figure}

Figure~\ref{fig:storm_architecture} depicts the main components of a Storm cluster \cite{AllenStormApplied:2015}. Storm uses a master-slave execution architecture where a \textit{Master Node}, which runs a daemon called \textit{Nimbus}, is responsible for scheduling tasks among \textit{Worker Nodes} and for maintaining a membership list to ensure reliable data processing. Nimbus interacts with Zookeeper \cite{ZooKeeper} to detect node failure and reassign tasks accordingly if needed. A Storm cluster comprises multiple worker nodes, each worker representing a virtual or physical machine. A worker node runs a  \textit{Supervisor} daemon, and one or multiple \textit{Worker Processes}, which are processes (\ie a JVM) spawned by Storm and able to run one or more \textit{Executors}. An executor thread executes one or more tasks. A \textit{Task} is both a realisation of a topology node and an abstraction of a Spout or Bolt. A Spout is a data stream source; it is the component responsible for reading the data from an external source and generating the data influx processed by the topology nodes. A Bolt listens to data, accepts a tuple, performs a computation or transformation -- \eg filtering, aggregation, joins, query databases, and other \acp{UDF} -- and optionally emits a new tuple.
 
Storm has many configuration options to define how topologies make use of host resources. An administrator can specify the number of worker processes that a node can create, also termed slots, as well as the amount of memory that slots can use. To parallelise nodes of a Storm topology a user needs to provide hints on how many concurrent tasks each topology component should run or how many executors to use; the latter influences how many threads will execute spouts and bolts. Tasks resulting from parallel Bolts perform the same function over different sets of data but may execute in different machines and receive data from different sources. Storm's scheduler, which is run by the Master, assigns tasks to workers in a round-robin fashion.

Storm allows for new worker nodes to be added to an existing cluster on which new topologies and tasks can be launched. It is also possible to modify the number of worker processes and executors spawned by each process. Modifying the level of parallelism by increasing or reducing the number of tasks that a running topology can create or the number of executors that it can use is more complex and, by default, requires the topology to be stopped and rebalanced. Such operation is expensive and can incur a considerable downtime. Moreover, some tasks may maintain state, perform grouping or hashing of tuple values that are henceforth assigned to specific downstream tasks. Stateful tasks complicate the dynamic adjustment of a running topology even further. As described in Section~\ref{sec:elasticity}, existing work has attempted to circumvent some of these limitations to enable resource elasticity.

Further performance tuning is possible by adjusting the length of executors' input and output queues, and worker processes' queues; factors that can impact the behaviour of the framework and its performance. Existing work has proposed changes to Storm to provide more predictable performance and hence meet some of the requirements of real time applications \cite{BasantaRealTime:2015}. By using Trident, Storm can also perform micro-batch processing. Trident topologies can be designed to act on batches of tuples that are grouped during short intervals and then processed by a task topology. Storm is also used by frameworks that provide high-level programming abstractions such as Summingbird \cite{BoykinSummingbird:2014} that mix multiple execution models.

\acrodef{HeronTM}[TM]{Topology Master}
\acrodef{HeronI}[HI]{Heron Instance}
\acrodef{HeronSM}[SM]{Stream Manager}
\acrodef{HeronMM}[MM]{Metrics Manager}
\acrodef{HeronT}[HT]{Heron Tracker}

\subsubsection{Twitter Heron} 
While maintaining API compatibility with Apache Storm, Twitter's Heron \cite{KulkarniHeron:2015} was built with a range of architectural improvements and mechanisms to achieve better efficiency and to address several of Storm issues highlighted in previous work \cite{ToshniwalStormAtTwitter:2014}. Heron topologies are process-based with each process running in isolation, which eases debugging, profiling, and troubleshooting. By using its built-in back pressure mechanisms, topologies can self-adjust when certain components lag.

Similarly to Storm, Heron topologies are directed graphs whose vertices are either \textit{Spouts} or \textit{Bolts} and edges represent streams of \textit{tuples}. The data model consists of a \textit{logical plan}, which is the description of the topology itself and is analogous to a database query; and the \textit{physical plan} that maps the actual execution logic of a topology to the physical infrastructure, including the machines that run each spout or bolt. When considering the execution model, Heron topologies comprise the following main components: \textit{Topology Master}, \textit{Container}, \textit{Stream Manager}, \textit{Heron Instance}, \textit{Metrics Manager}, and \textit{Heron Tracker}.

\begin{figure}[ht]
\centering 
\includegraphics[width=1.\linewidth]{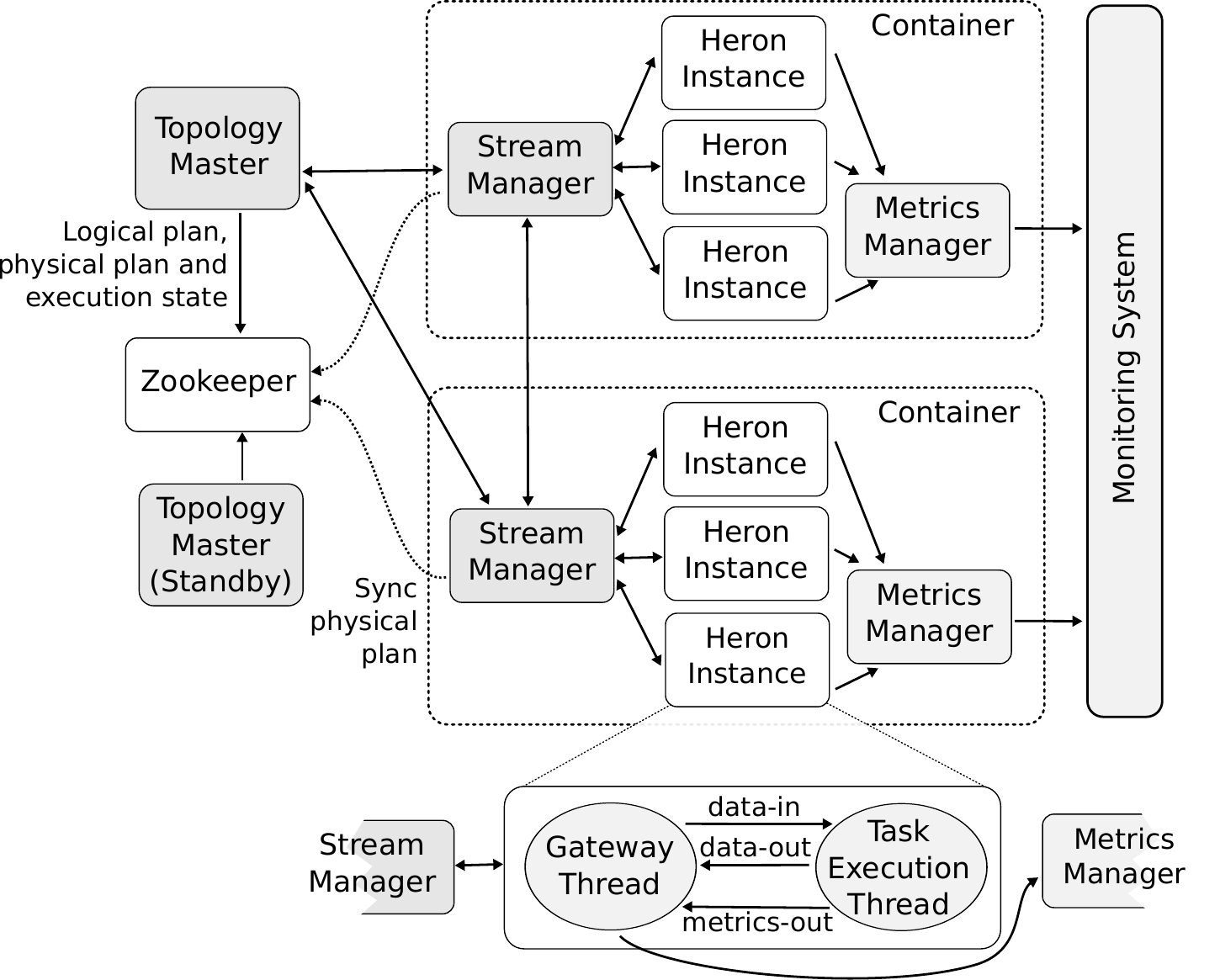} 
\caption{Main architecture components of a Heron topology \cite{KulkarniHeron:2015}.}
\label{fig:heron_topology}
\end{figure}

Heron provides a command-line tool for submitting topologies to the \textit{Aurora Scheduler}, a scheduler built to run atop Mesos \cite{HindmanMesos:2011}. Heron can also work with other schedulers including YARN, and Amazon \ac{AmazonECS} \cite{AmazonECS}. Support to other schedulers is enabled by an abstraction designed to avoid the complexity of Storm Nimbus, often highlighted as an architecture issue in Storm. A topology in Heron runs as an Aurora job that comprises multiple \textit{Containers}. 

When a topology is deployed, Heron starts a single \textit{\ac{HeronTM}} and multiple containers (Figure~\ref{fig:heron_topology}). The \ac{HeronTM} manages the topology throughout its entire life cycle until a user deactivates it. Zookeeper \cite{ZooKeeper} is used to guarantee that there is a single \ac{HeronTM} for the topology and that it is discoverable by other processes. The \ac{HeronTM} also builds the physical plan and serves as a gateway for topology metrics. Heron allows for creating a StandBy \ac{HeronTM} in case the main \ac{HeronTM} fails.
Containers communicate with the \ac{HeronTM} hence forming a fully connected graph. Each container hosts multiple \textit{\acp{HeronI}}, a \textit{\ac{HeronSM}}, and a \textit{\ac{HeronMM}}. An \ac{HeronSM} manages the routing of tuples, whereas \acp{HeronSM} in a topology form a fully connected network.
Each \ac{HeronI} communicates with its local \ac{HeronSM} when sending and receiving tuples. The work for a spout and a bolt is carried out by \acp{HeronI}, which unlike Storm workers, are JVM processes. 
An \ac{HeronMM} gathers performance metrics from components in a container, which are in turn routed both to the \ac{HeronTM} and external collectors. An \textit{\ac{HeronT}} is a gateway for cluster-wide information about topologies.

An \ac{HeronI} follows a two-threaded design with one thread responsible for executing the logic programmed as a spout or bolt (\ie Execution), and another thread for communicating with other components and carrying out data movement in and out of the \ac{HeronI} (\ie Gateway). The two threads communicate with one another via three unidirectional queues, of which two are used by the Gateway to send/receive tuples to/from the Execution thread, and another is employed by the Execution thread to export collected performance metrics.

\acrodef{S4PN}[PN]{Processing Node}

\subsubsection{Apache \acs{S4}} 
The \acf{S4} \cite{NeumeyerS4:2010} is a distributed stream processing engine that uses the actor model for managing concurrency. \acp{PE} perform computation and exchange events, where each \ac{PE} can handle data events and either emit new events or publish results. 

\begin{figure}[ht]
\centering 
\includegraphics[width=.9\linewidth]{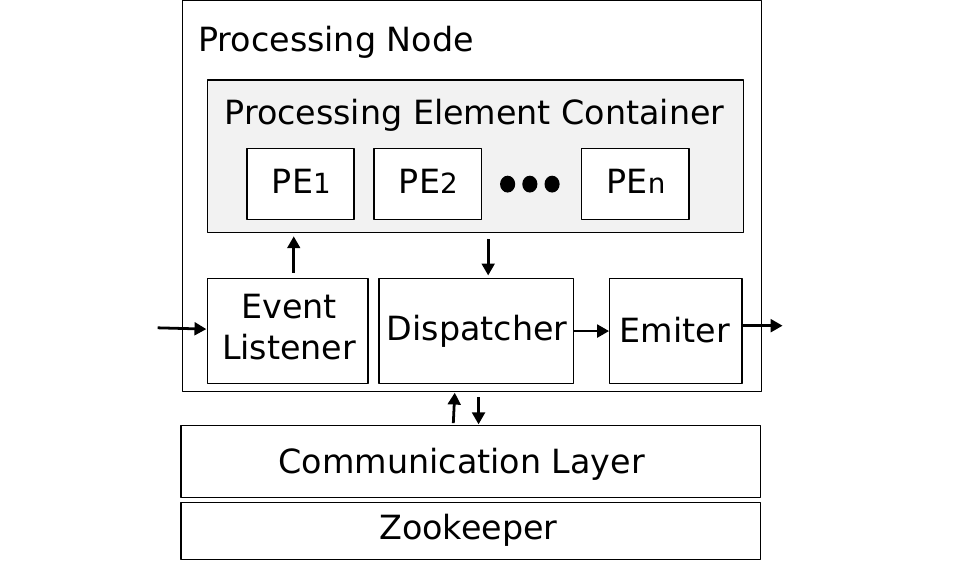} 
\caption{A processing node in \ac{S4} \cite{NeumeyerS4:2010}.}
\label{fig:s4_pe}
\end{figure}

\ac{S4} can use commodity cluster hardware and employs a decentralised and symmetric runtime architecture comprising \acp{S4PN} that are homogeneous concerning functionality. As depicted in Figure \ref{fig:s4_pe}, a \ac{S4PN} is a machine that hosts a container of \acp{PE} that receive events, execute user-specified functions over the data, and use the communication layer to dispatch and emit new events. ZooKeeper \cite{ZooKeeper} provides features used for coordination between \acp{S4PN}. 

When developing a \ac{PE}, a developer must specify its functionality and the type of events it can consume. While most \acp{PE} can only handle events with given keyed attribute values, \ac{S4} provides a keyless \ac{PE} used by its input layer to handle all events that it receives. \acp{S4PN} route events using a hash function of their keyed attribute values. Following receipt of an event, a listener passes it to the processing element container that in turn delivers it to the appropriate \acp{PE}.

\subsubsection{Apache Samza} 

Apache Samza \cite{ApacheSamza} is a stream processing framework that uses Apache Kafka for messaging and Apache YARN \cite{VavilapalliYARN:2013} for deployment, resource management, and security. A Samza application is a data flow that consists of \textit{consumers} that fetch data events that processed by a graph of jobs, each job containing one or multiple tasks. Unlike Storm, however, where topologies need to be deployed as a whole, Samza does not natively support the \ac{DAG} topologies. In Samza, each job is an entity that can be deployed, started or stopped independently. 

Like Heron, Samza uses single-threaded processes (containers), mapped to one CPU core. Each Samza task contains an embedded key-value store used to record state. Changes to this key-value store are replicated to other machines in the cluster allowing for tasks to be restored quickly in case of failure.

\subsubsection{Apache Flink}

Flink offers a common runtime for data streaming and batch processing applications \cite{ApacheFlink}. Applications are structured as arbitrary \acp{DAG}, where special cycles are enabled via iteration constructs. Flink works with the notion of \textit{streams} onto which \textit{transformations} are performed. A stream is an intermediate result, whereas a transformation is an operation that takes one or more streams as input, and computes one or multiple streams. During execution, a Flink application is mapped to a \textit{streaming workflow} that starts with one or more \textit{sources}, comprises \textit{transformation operators}, and ends with one or multiple \textit{sinks}. Although there is often a mapping of one transformation to one dataflow operator, under certain cases, a transformation can result in multiple operators. Flink also provides APIs for iterative graph processing, such as Gelly \cite{FlinkGelly}.

The parallelism of Flink applications is determined by the degree of parallelism of streams and individual operators. Streams can be divided into \textit{stream partitions} whereas operators are split into \textit{subtasks}. Operator subtasks are executed independently from one another in different threads that may be allocated to different containers or machines.

\begin{figure}[ht]
\centering 
\includegraphics[width=1.\linewidth]{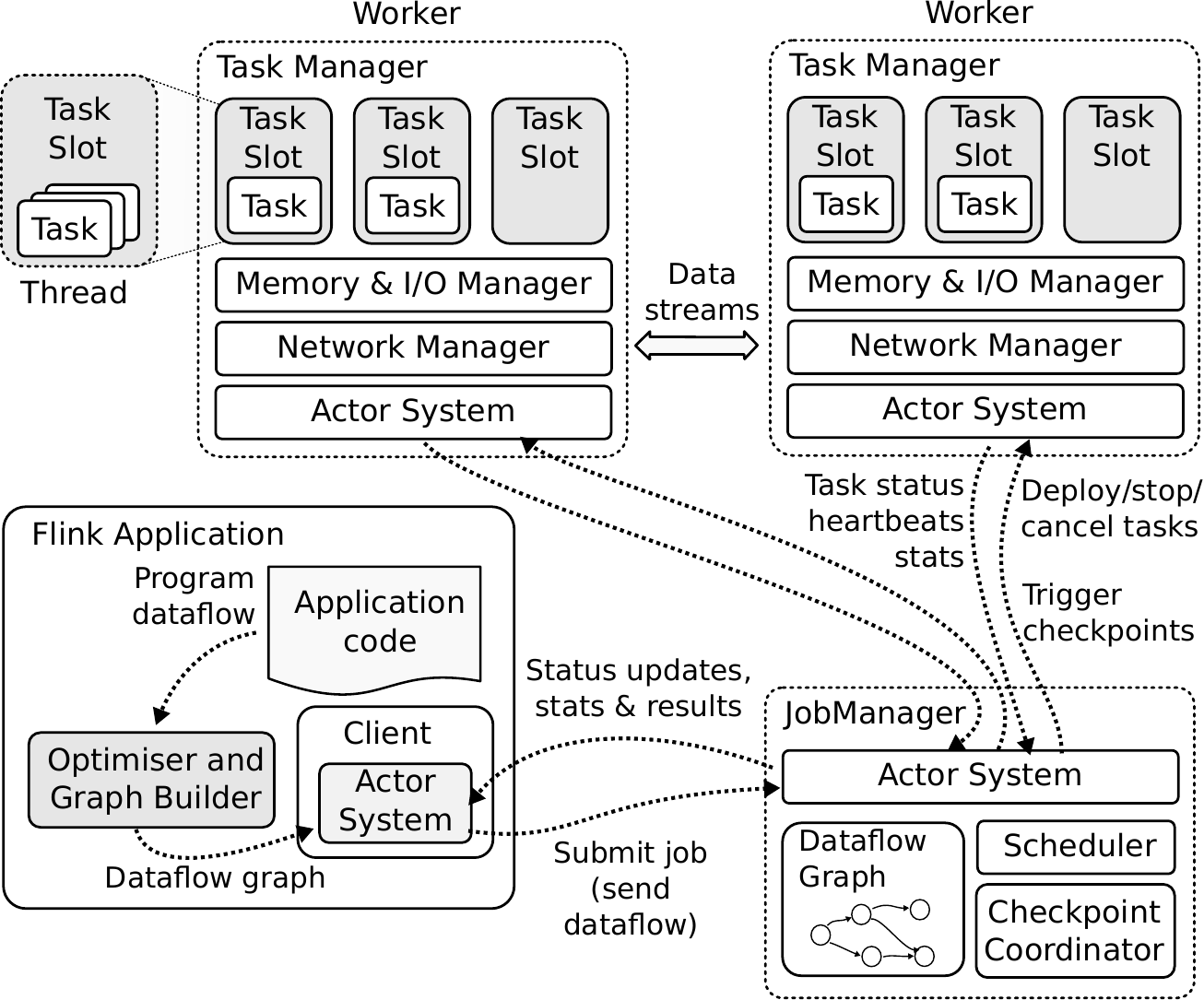} 
\caption{Apache Flink's execution model \cite{ApacheFlink}.}
\label{fig:flink_execution}
\end{figure}

Flink's execution model (Figure \ref{fig:flink_execution}) comprises two types of processes, namely a master also called the \textit{JobManager} and workers termed as \textit{TaskManagers}. The JobManager is responsible for coordinating the scheduling tasks, checkpoints, failure recovery, among other functions. TaskManagers execute subtasks of a Flink dataflow. They also buffer and exchange data streams. A user can submit an application using the Flink client, which prepares and sends the dataflow to a JobManager.

Similar to Storm, a Flink worker is a JVM process that can execute one or more subtasks in separate threads. The worker also uses the concept of slots to configure how many execution threads can be created. Unlike Storm, Flink implements its memory management mechanism that enables a fair share of memory that is dedicated to each slot.    

\subsubsection{Spark Streaming} 

Apache Spark is a cluster computing solution that extends the MapReduce model to support other types of computations such as interactive queries and stream processing \cite{ZahariaRDD:2012}. Designed to cover a variety of workloads, Spark introduces an abstraction called \acp{RDD} that enables running computations in memory in a fault-tolerant manner. \acp{RDD}, which are immutable and partitioned collections of records, provide a programming interface for performing operations, such as map, filter and join, over multiple data items. For fault-tolerance purposes, Spark records all transformations carried out to build a dataset, thus forming the so-called \textit{lineage graph}.

\begin{figure}[ht]
\centering 
\includegraphics[width=.95\linewidth]{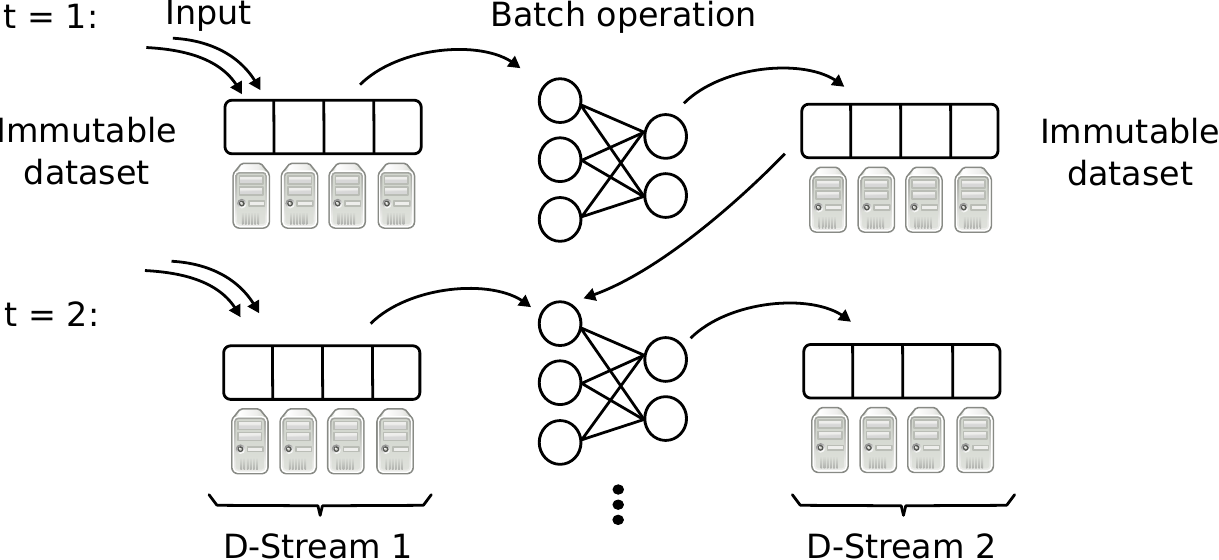} 
\caption{D-Stream processing model \cite{ZahariaDiscretizedStreams:2013}.}
\label{fig:dstreams}
\end{figure}

Under the traditional stream processing approach based on a graph of continuous operators that process tuples as they arrive, it is arguably difficult to achieve fault tolerance and handle stragglers. As application state is often kept by multiple operators, fault tolerance is achieved either by replicating sections of the processing graph or via upstream backup. The former demands synchronisation of operators via a protocol such as Flux \cite{ShaFlux:2003} or other transactional protocols \cite{Wu:2015}, whereas the latter, when a node fails, requires parents to replay previously sent messages to rebuild the state. 

To handle faults and stragglers more efficiently, Zaharia \etal \cite{ZahariaDiscretizedStreams:2013} proposed D-Streams, a discretised stream processing based on Spark Streaming. 
As depicted in Figure \ref{fig:dstreams}, D-Streams follows a micro-batch approach that organises stream processing as batch computations carried out periodically over small time windows. During a short time interval, D-Streams stores the received data, which the cluster resources then use as input dataset for performing parallel computations once the interval elapses. These computations produce new datasets that represent an intermediate state or computation outputs. The intermediate state consists of \acp{RDD} that D-Streams processes along with the datasets stored during the next interval. In addition to providing a strong unification with batch processing, this model stores the state in memory as \acp{RDD} \cite{ZahariaRDD:2012} that D-Streams can deterministically recompute.

\subsubsection{Other Solutions}
System S, a precursor to IBM Streams\footnote{IBM has rebranded its data stream processing solution a few times over the years. Although some papers mention System S and InfoSphere Streams, hereafter we employ simply \textit{IBM Streams} to refer to IBM's stream processing solution.}, is a middleware that organises applications as \acp{DAG} of operators and that supports distributed processing of both structured and unstructured data streams. \ac{SPL} offers a language and engine for composing distributed and parallel data-flow graphs and a toolkit for building generic operators \cite{HirzelSPL:2017}. It provides language constructs and compiler optimisations that utilise the performance of the \ac{SPC} \cite{AminiSPC:2006}. \ac{SPC} is a system for designing and deploying stream processing \acp{DAG} that support both relational operators and user-defined operators. It places operators on containers that consist of processes running on cluster nodes. The \ac{SPC} data fabric provides the communication substrate implemented on top of a collection of distributed servers.

\textsc{Esc} \cite{SatzgerESC:2011} is another stream processing engine that also follows the data-flow scheme where programs are \acp{DAG} whose vertices represent operations performed on the received data and edges are the composition of operators. The \textsc{Esc} system, which uses the actor model for concurrency, comprises a system and multiple machine processes responsible for executing workers.

Other systems, such as TimeStream \cite{QianTimeStream:2013}, use a \ac{DAG} abstraction for structuring an application as a graph of operators that execute user-defined functions. Employing a graph abstraction is not exclusive to data stream processing. Other big data processing frameworks \cite{SahaTez:2015} also provide high-level APIs that enable developers to specify computations as a \ac{DAG}. The deployment of such computations is performed by engines using resource management systems such as Apache YARN.

Google's MillWheel \cite{AkidauMillWheel:2013} also employs a data flow abstraction in which users specify a graph of transformations, or computations, that are performed on input data to produce output data. MillWheel applications run on a dynamic set of hosts where each computation can run on one or more machines. A master node manages load distribution and balancing by dividing each computation into a set of key intervals. Resource utilisation is continuously measured to determine increased pressure, in which case intervals are moved, split, or merged.

The \ac{ELF} stream processing system \cite{HuELF:2014} uses a decentralised architecture with `in-situ' data access where each job extracts data directly from a Web server, placing it in compressed buffer trees for local parsing and temporary storage. The data is subsequently aggregated using shared reducer trees mapped to a set of worker processes executed by agents structured as an overlay built using Pastry \ac{DHT}. \ac{ELF} attempts to overcome some of the limitations of existing solutions that require data movement across machines and where the data must be somewhat stale before it arrives at the stream processing system. 


%% file: cloud_systems.tex
\section{Managed Cloud Systems}
\label{sec:cloud_systems}

This section describes public cloud solutions for processing streaming data and presents details on how elasticity features are made available to developers and end users. The section primarily identifies prominent technological solutions for processing of streaming data and highlights their main features.

\subsection{\ac{AWS} Kinesis}
A streaming data service can use Firehose for delivering data to \ac{AWS} services such as Amazon Redshift, Amazon \ac{S3}, or Amazon \ac{ES}. It works with data producers or agents that send data to Firehose, which in turn delivers the data to the user-specified destination or service. When choosing \ac{S3} as the destination, Firehose copies the data to an \ac{S3} bucket. Under Redshift, Firehose first copies the data to an \ac{S3} bucket before notifying Redshift. Firehose can also deliver the streaming data to an \ac{ES} cluster.

Firehose works with the notion of \textit{delivery streams} to which \textit{data producers} or agents can send data \textit{records} of up to 1000 KB in size. Firehose buffers incoming data up to a \textit{buffer size} or for a given \textit{buffer interval} in seconds before it delivers the data to the destination service. Integration with the Amazon CloudWatch \cite{CloudWatch} enables monitoring the number of bytes transferred, the number of records, the success rate of operations, time taken to perform certain operations on delivery streams, among others. \ac{AWS} enforces certain limits on the rate of bytes, records and number of operations per delivery stream, as well as streams per region and \ac{AWS} account.

Amazon Kinesis Streams is a service that enables continuous data intake and processing for several types of applications such as data analytics and reporting, infrastructure log processing, and complex event processing. Under Kinesis Streams \textit{producers} continuously push data to Streams, which is then processed by \textit{consumers}. A \textit{stream} is an ordered sequence of \textit{data records} that are distributed into \textit{shards}. A Kinesis Streams \textit{application} is a consumer of a stream that runs on Amazon \ac{EC2}. A shard has a fixed data capacity regarding reading operations and the amount of data read per second. The total capacity of a stream is the aggregate capacity of all of its shards. Integration with Amazon CloudWatch allows for monitoring the performance of the available streams. A user can adjust the capacity of a stream by \textit{resharding} it. Two operations are allowed for respectively increasing or decreasing available capacity, namely splitting an existing shard or merging two shards.      

\subsection{Google Dataflow}

Google Cloud Dataflow \cite{GoogleDataFlow} is a programming model and managed service for developing and executing a variety of data processing patterns such as \ac{ETL} tasks, batch processing, and continuous computing.

Dataflow's programming model enables a developer to specify a data processing job that is executed by the Cloud Dataflow runner service. A data processing job is specified as a \textit{Pipeline} that consists of a directed graph of steps or \textit{Transforms}. A transform takes one or more \textit{PCollection}'s -- that represent data sets in the pipeline -- as input, performs the user-provided processing function on the elements of the PCollection and produces an output PCollection. A PCollection can hold data of a fixed size, or an unbounded data set from a continuously updating source. For unbounded sources, Dataflow enables the concept of \textit{Windowing} where elements of the PCollection are grouped according to their timestamps. A \textit{Trigger} can be specified to determine when to emit the aggregate results of each window. Data can be loaded into a Pipeline from various \textit{I/O Sources} by using the Dataflow SDKs as well as written to output \textit{Sinks} using the sink APIs. As of writing, the Dataflow SDKs are being open sourced under the Apache Beam incubator project \cite{ApacheBeam}.

The Cloud Dataflow managed service can be used to deploy and execute a pipeline. During deployment, the managed service creates an execution graph, and once deployed the pipeline becomes a Dataflow job. The Dataflow service manages services such as Google Compute Engine \cite{GoogleComputeEngine} and Google Cloud Storage \cite{GoogleCloudStorage} to run a job, allocating and releasing the necessary resources. The performance and execution details of the job are made available via the Monitoring Interface or using a command-line tool. The Dataflow service attempts to perform certain automatic job optimisations such as data partitioning and parallelisation of worker code, optimisations of aggregation operations or fusing transforms in the execution graph. 

On-the-fly adjustment of resource allocation and data partitioning are also possible via Autoscaling and Dynamic Work Rebalancing. For bounded data in batch mode Dataflow chooses the number of VMs based on both the amount of work in each step of a pipeline and the current throughput. Although autoscaling can be used by any batch pipeline, as of writing autoscaling for streaming-mode is experimental and participation is restricted to invited developers. It is possible, however, to adjust the number of workers assigned to a streaming pipeline manually, which replaces a running job with a new job while preserving the state information.

\acrodef{SU}{Streaming Unit}
\subsection{Azure Stream Analytics}

\ac{ASA} enables real-time analysis of streaming data from several sources such as devices, sensors, websites, social media, applications, infrastructures, among other sources \cite{AzureStreamAnalytics}.  

A job definition in \ac{ASA} comprises data \textit{inputs}, a \textit{query}, and data \textit{output}. Input is the data streaming source from which the job reads the data, a query transforms the received data, and the output is to where the job sends results. Stream Analytics provides integration with multiple services and can ingest streaming data from Azure Event Hubs and Azure IoT Hub, and historical data from Azure Blob service. It performs analytic computations that are specified in a declarative language; a T-SQL variant termed as Stream Analytics Query Language. Results from Stream Analytics can be written to several data sinks such as Azure Storage Blobs or Tables, Azure SQL DB, Event Hubs, Azure Service Queues, among other sinks. They can also be visualised or further processed using other tools deployed on Azure compute cloud. As of writing, Stream Analytics does not support \acp{UDF} for data transformation.

The allocation of processing power and resource capacity to a Stream Analytics job is performed considering \acp{SU} where an \ac{SU} represents a blend of CPU capacity, memory, and read/write data rates.  Certain query steps can be partitioned, and some \acp{SU} can be allocated to process data from each partition, hence increasing throughput. To enable partitioning the input data source must be partitioned and the query modified to read from a partitioned data source. 


%% file: elasticity.tex
\acrodef{SC}{StreamCloud}
\acrodef{BIM}{Bucket-Instance Map}
\acrodef{LB}{Load Balancer}
\acrodef{EM}{Elastic Manager}
\acrodef{RM}{Resource Manager}
\acrodef{LM}{Local Manager}
\acrodef{MCEP}{Mobile Complex Event Processing}

\section{Elasticity in Stream Processing Systems}
\label{sec:elasticity}

Over time several types of applications have benefited from resource elasticity, a key feature of cloud computing \cite{LoridoBotranElasticity:2014}. As highlighted by Lorido-Botran \etal, elasticity in cloud environments is often accomplished via a \textit{\ac{MAPE}} process where:

\begin{enumerate}
  \item application and system metrics are \textit{monitored}; 
  \item the gathered information is \textit{analysed} to assess current performance and utilisation, and optionally predict future load;
  \item based on an auto-scaling policy an auto-scaler creates an elasticity \textit{plan} on how to add or remove capacity; and
  \item the plan is finally \textit{executed}.
\end{enumerate}

After analysing performance data, an auto-scaler may choose to adjust the number of resources (\eg add or remove compute resources) available to running, newly submitted, applications. Managing elasticity of data stream processing applications often requires solving two inter-related problems: (i) allocating or releasing IT resources to match an application workload; and (ii) devising and performing actions to adjust the application to make use of the additional capacity or release previously allocated resources. The first problem, which consists in modifying the resource pool available for a stream processing application, is termed here as \textit{elastic resource management}. A decision made by a resource manager to add/remove resource capacity for a stream processing application is referred to as \textit{scale out/in plan}\footnote{The term scale out/in is often employed in horizontal elasticity, but a plan can also be scale up/down when using vertical elasticity. For brevity, we use only scale out/in in the rest of the text}. We refer to the actions taken to adjust an application during a scale out/in plan as \textit{elasticity actions}. 

\begin{figure}[ht]
\centering 
\includegraphics[width=1.\linewidth]{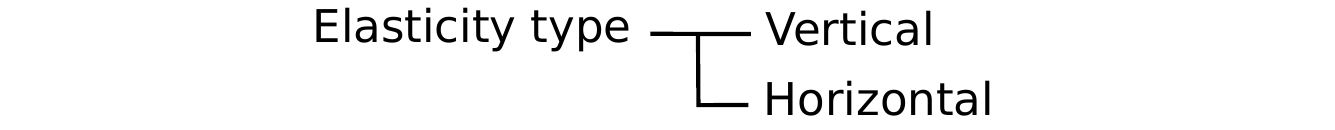} 
\caption{Types of elasticity used by elastic resource management.}
\label{fig:elasticity_types}
\end{figure}

Similarly to other services running in the cloud, elastic resource management for data stream processing applications can make use of two types of elasticity, namely \textit{vertical} and \textit{horizontal} (Figure~\ref{fig:elasticity_types}), which have their impact on the kind of elastic actions for adapting an application. Vertical elasticity consists in allocating more resources such as CPU, memory and network capacity on a host that has previously been allocated to a given application. As described later, stream processing can benefit from this type of elasticity by, for instance, increasing the instances of a given operator (\ie operator \textit{fission} \cite{HirzelStreamOptimisation:2014}). Horizontal elasticity consists essentially in allocating additional computing nodes to host a running application. 

To make use of additional resources and improve application performance, auto-scaling operations may require adjusting applications dynamically by, for example, performing optimisations in their execution graphs, or modifying intra-query parallelism by increasing the number of instances of certain operators. Previous work has discussed approaches on reconfiguration schemes to modify the placement of stream processing operators dynamically to adjust an application to current resource conditions or provide fault-tolerance \cite{LakshmananPlaceStrategies:2008}. The literature on data stream processing often employs the term \textit{elastic} to convey \textit{operator placement schemes} that enable applications to deliver steady performance as their workload increases, not necessarily exploring the types of elasticity mentioned above.

Although the execution of scale out/in plans presents similarities with other application scenarios (\eg adding/removing resources from a resource pool), adjusting a stream processing system and applications dynamically to make use of the newly available capacity or release unused resources is not a trivial task. The enforcement of scale out/in plans faces multiple challenges. Horizontal elasticity often requires adapting the graph of processing elements and protocols, exporting and saving operator state for replication, fault tolerance and migration. As highlighted by Sattler and Beier \cite{SattlerElastic:2013}, performing parallel processing is often difficult in the case of window- or sequence-based operators including \ac{CEP} operators due to the amount of state they keep. Elastic operations, such as adding nodes or removing unused capacity, may require at least re-routing the data, changing the manner an incoming dataflow is split among parallel processing elements, among other issues. Such adjustments are costly to perform, particularly if processing elements maintain state. As stream processing queries are often treated as long running that cannot be restarted without incurring a loss of data, the initial operator placement (also called task assignment), where processing elements are deployed on available computing resources becomes more critical than in other systems.

\begin{figure}[ht]
\centering 
\includegraphics[width=1.\linewidth]{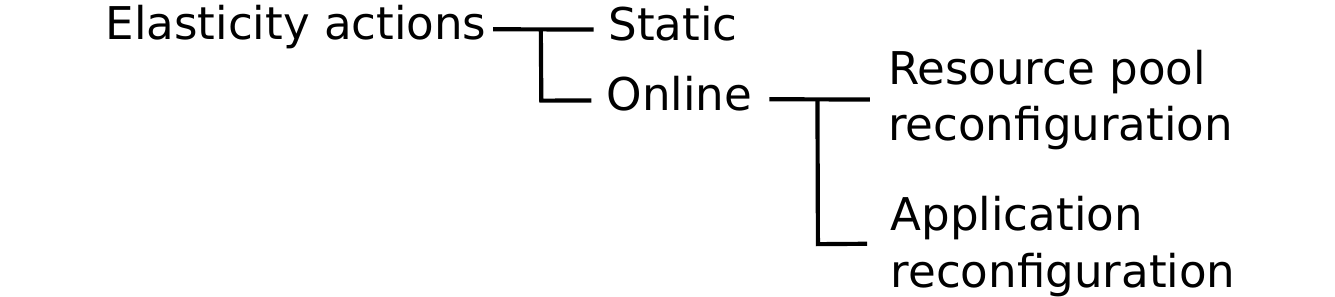} 
\caption{Elasticity actions for stream processing engines.}
\label{fig:dataflow_elasticity}
\end{figure}

Given how important the initial task assignment is to guarantee the elasticity of stream processing systems, we classify elasticity actions into two main categories, namely \textit{static} and \textit{online} as depicted in Figure~\ref{fig:dataflow_elasticity}. When considering the operator \ac{DAG} based solutions, static techniques comprise optimisations made to modify the original graph (\ie the logical plan) to improve task parallelism and operator placement, optimise data transfers, among other goals \cite{HirzelStreamOptimisation:2014}. Previous work provided a survey of various static techniques \cite{LakshmananPlaceStrategies:2008}. Online approaches comprise both actions to modify the pool of available resources and dynamic optimisations carried out to adjust applications dynamically to utilise newly allocated resources. The next sections provide more details on how existing solutions address challenges in these categories with a focus on online techniques.


\subsection{Static Techniques}

A review of strategies for placing processing operators in early distributed data stream processing systems has been presented in previous work \cite{LakshmananPlaceStrategies:2008}. Several approaches for optimising the initial task assignment or scheduling exploit intra-query parallelism by ensuring that certain operators can scale horizontally to support larger numbers of incoming tuples, thus achieving greater throughput. 

R-Storm \cite{PengRStorm:2015} handles the problem of task assignment in Apache Storm by providing custom resource-aware scheduling schemes. Under the considered approach, each task in a Storm topology has soft CPU and bandwidth requirements and a hard memory requirement. The available cluster nodes, on the other hand, have budgets for CPU, bandwidth and memory. While considering the throughput contribution of a data sink, given by the rate of tuples it is processing, R-Storm aims to assign tasks to a set of nodes that increases overall throughput, maximises resource utilisation, and respects resource budgets. The assignment scenario results is a quadratic multiple 3-dimensional knapsack problem. After reviewing existing solutions with several variants of knapsack problems, the authors concluded that existing methods are computationally expensive for distributed stream processing scenarios. They proposed scheduling algorithms that view a task as a vector of resource requirements and nodes as vectors of resource budgets. The algorithm uses the Euclidean distance between a task vector and node vectors to select a node to execute a task. It also uses heuristics that attempt to place tasks that communicate in proximity to one another, that respect hard constraints, and that minimise resource waste.

Pietzuch \etal \cite{PietzuchOpPlaceNetwork:2006} create a \ac{SBON} between a stream processing engine and the physical network. \ac{SBON} manages operator placement while taking into account network latency. The system architecture uses an adaptive optimisation technique that creates a multidimensional Euclidean space, termed as the \textit{cost space}, over which the placement is projected. Optimisation techniques such as spring relaxation are used to compute operator placement using this mathematical space. A proposed scheme maps a solution obtained using the cost space onto physical nodes.

The scheme proposed by Zhou \etal also \cite{ZhouOpPlacement:2006} for the initial operator placement attempts to minimise the communication cost whereas the dynamic approach considers load balancing of scheduled tasks among available resources. The initial placement schemes group operators of a query tree into query fragments and try to minimise the number of compute nodes to which they are assigned. Ahmad and \c{C}etintemel \cite{AhmadOpPlacement:2004} also proposed algorithms for the initial placement of operators while minimising the bandwidth utilised in the network, even though it is assumed that the algorithms could be applied periodically.

\subsection{Online Techniques}

Systems for providing elastic stream processing on the cloud generally comprise two key elements: 

\begin{itemize}
\item a subsystem that monitors how the stream processing system is utilising the available resources (\eg use of CPU, memory and network resources) \cite{FernandezScaleOut:2013} and/or other service-level metrics (\eg number of tuples processed over time, tail end-to-end latency \cite{HeinzeFUGU:2014}, critical paths \cite{ViglasRateOptimisation:2002}) and tries to identify bottleneck operators; and
\item a scaling policy that determines when scale out/in plans should be performed \cite{LohrmannLatency:2015}.
\end{itemize}

As mentioned earlier, in addition to adding/removing resources, a scale out/in plan is backed by mechanisms to adjust the query graph to make efficient use of the updated resource pool. Proposed mechanisms consist of, for instance, increasing operator parallelism; rewriting the query graph based on certain patterns that are empirically proven to improve performance and rewriting rules specified by the end user; and migrating operators to less utilised resources.

Most solutions are application and workload agnostic -- \ie do not attempt to model application behaviour or detect changes in the incoming workload \cite{KrishnamurthySketchChange:2003} -- and offer methods to: (i) optimise the initial scheduling, when processing tasks are assigned to and deployed onto available resources; and/or (ii) reschedule processing tasks dynamically to take advantage of an updated resource pool. Operators are treated as black boxes and (re)scheduling and elastic decisions are often taken considering a performance metric. Certain solutions that are not application-agnostic attempt to identify workload busts and behaviours by considering characteristics of the incoming data as briefly described in Section~\ref{subsec:change_detection}.

Sattler and Beier \cite{SattlerElastic:2013} argue that distributing query nodes or operators can improve reliability \textit{``by introducing redundancy, and increasing performance and/or scalability by load distribution''}. They identify operator patterns -- \eg simple standby, check-pointing, hot standby, stream partitioning and pipelining -- for building rules for restructuring the physical plan of an application graph, which can increase fault tolerance and achieve elasticity. They advocate that re-writings should be performed when a task becomes a bottleneck; \ie it cannot keep up with the rate of incoming tuples. An existing method is used to scan the execution graph and find critical paths based on monitoring information gathered during query execution \cite{ViglasRateOptimisation:2002}.  

While dynamically adjusting queries with stateless operators can be difficult, modifying a graph of stateful operators to increase intra-query parallelism is more complex. As stated by Fernandez \etal \cite{FernandezScaleOut:2013}, during adjustment, operator \textit{``state must be partitioned correctly across a larger set of VMs''}. Fernandez \etal hence propose a solution to manage operator state, which they integrate into a stream processing engine to provide \textit{scale out} features. The solution offers primitives to export operator state as a set of tuples, which is periodically check-pointed by the processing system. An operator keeps state regarding its processing, buffer contents, and routeing table. During a scale out operation, the key space of the tuples that an operator handles is repartitioned, and its processing state is split across the new operators. The system measures CPU utilisation periodically to detect bottleneck operators. If multiple measurements are above a given threshold, then the scale-out coordinator increases the operator parallelism.

Previous work has also attempted to improve the assignment of tasks and executors to available resources in Storm and to reassign them dynamically at runtime according to resource usage conditions. T-Storm \cite{XuTStorm:2014} (\ie Traffic-aware Storm), for instance, aims to reduce inter-process and inter-node communication, which is shown to degrade performance under certain workloads. T-Storm monitors workload and traffic load during runtime. It provides a scheduler that generates a task schedule periodically, and a custom Storm scheduler that fetches the schedule and executes it by assigning executors accordingly. Aniello \etal provide a similar approach, with two custom Storm schedulers, one for offline static task assignment and another for dynamic scheduling \cite{AnielloStormAdaptive:2013}. Performance monitoring components are also introduced, and the proposed schedulers aim to reduce inter-process and inter-node communication.

Lohrmann \etal \cite{LohrmannLatency:2015} introduced policies that use application or system performance metrics such as CPU utilisation thresholds, the rate of tuples processed per operator, and tail end-to-end latency. They propose a strategy to provide latency guarantees in stream processing systems that execute heady \ac{UDF} data flows while aiming to minimise resource utilisation. The reactive strategy (\ie \textit{ScaleReactively}) aims to enforce latency requirements under varying load conditions without permanently overprovisioning resource capacity. The proposed solution assumes homogeneous cluster nodes, effective load balancing of elements executing \acp{UDF}, and elastically scalable \acp{UDF}. The system architecture comprises elements for monitoring the latency incurred by operators in a job sequence. The reactive strategy uses two techniques, namely \textit{Rebalance} and \textit{ResolveBottlenecks}. The former adjusts the parallelism of bottleneck operators whereas the latter, as the name implies, resolves bottlenecks by scaling out so that the first technique can be applied again at later time.

The \textsc{Esc} stream processing system \cite{SatzgerESC:2011} comprises several components for task scheduling, performance monitoring, management of a resource pool to/from which machines are added/released, as well as application adaptation decisions. A processing element process executes \acp{UDF} and contains a manager and multiple workers, which serve respectively as a gateway for the element itself and for executing multiple instances of the \ac{UDF}. The \ac{PE} manager employs a function for balancing the load among workers. Each worker contains a buffer or queue and an operator. The autonomic manager of the system process monitors the load of machines and the length of the worker processes. For adaptation purposes, the autonomic manager can add/remove machines, replace the load balancing function of a \ac{PE} manager and spawn/kill new workers, kill the \ac{PE} manager and its workers altogether. The proposed elastic policies are based on load thresholds that, when exceeded, trigger the execution of actions such as attaching new machines.

\ac{SC} \cite{GulisanoStreamCloud:2012} provides multiple cloud parallelisation techniques for splitting stream processing queries that it assigns to independent subclusters of computing resources. According to the chosen technique, the number of resulting subqueries depends on the number of stateful operators that the original query contains. A subquery comprises a stateful operator and all intermediate stateless operators until another stateful operator or a data sink. \ac{SC} also introduces \textit{buckets} that receive output tuples from a subcluster. \acp{BIM} control the distribution of buckets to downstream subclusters, which may be dynamically modified by \acp{LB}. A load balancer is an operator that distributes tuples from a subquery to downstream subqueries. To manage elasticity, \ac{SC} employs a resource management architecture that monitors CPU utilisation and, if the utilisation is out of pre-determined lower or upper thresholds, it can: adjusts the system to rebalance the load; or provision or releases resources. 


Heinze \etal \cite{HeinzeFUGU:2014} attempt to model the spikes in a query's end-to-end latency when moving operators across machines, while trying to reduce the number of latency violations. Their target system, FUGU, considers two classes of scaling decisions, namely mandatory, which are operator movements to avoid overload; and optional, such as releasing an unused host during light load. FUGU employs the Flux protocol for migrating stream processing operators \cite{ShaFlux:2003}. Algorithms are proposed for scale out/in operations as well as operator placement. The scale-out solution extends the subset sum algorithm, where subsets of operators whose total load is below a pre-established threshold are considered to remain in a host. To pick a final set, the algorithm takes into consideration the latency spikes caused by moving the operators that are not in the set. For scale-in, FUGU releases a host with minimum latency spike. The operator placement is an incremental bin packing problem, where bins are nodes with CPU capacity, and items are operators with CPU load as weight.  Memory and network are second-level constraints that prevent placing operators on overloaded hosts. A solution based on the \textit{FirstFit} decreasing heuristic is provided.

Gedik \etal \cite{GedikElastic:2014} tackle the challenge of auto-parallelising distributed stream processing engines in general while focusing on IBM Streams. As defined by Gedik \etal \cite{GedikElastic:2014}, \textit{``auto-parallelisation involves locating regions in the application's data flow graph that can be replicated at run-time to apply data partitioning, in order to achieve scale.''} Their work proposes an elastic auto-parallelisation approach that handles stateful operators and general purpose applications. 
It also provides a control algorithm that uses metrics such as the blocking time at the splitter and throughput to determine how many parallel channels provide the best throughput. Data splitting for a parallel region can be performed in a round-robin manner if the region is stateless, or using a hash-based scheme otherwise. 

Also considering IBM Streams, Tang and Gedik \cite{TangAutoPipelining:2013} address task and pipeline parallelism by determining points of a data flow graph where adding additional threads can level out the resource utilisation and improve throughput. They consider an execution model that comprises a set of threads, where each thread executes a pipeline whose length extends from a starting operator port to a data sink or the port of another thread's first operator. They use the notion of utility to model the goodness of including a new thread and propose an optimisation algorithm find and evaluating parallelisation options. Gedik \etal \cite{GedikPipelinedFission:2016} propose a solution for IBM Streams exploiting pipeline parallelism and data parallelism simultaneously. They propose a technique that segments a chain-like data flow graph into regions according to whether the operators they contain can be replicated or not. For the parallelisable regions, replicated pipelines are created preceded and followed by, respectively split and merge operators.

Wu and Tan \cite{Wu:2015} discuss technical challenges that may require a redesign of distributed stream processing systems, such as maintaining large amounts of state, workload fluctuation and multi-tenant resource sharing. They introduce \textit{ChronoStream}, a system to support elasticity and high availability in latency-sensitive stream computing. To facilitate elasticity and operator migration, \textit{ChronoStream} divides the application-level state into a collection of \textit{computation slices} that are periodically check-pointed and replicated to multiple specified computing nodes using locality-sensitive techniques. In the case of component failure or workload redistribution, it reconstructs and reschedules slice computation. Unlike D-Streams, \textit{ChronoStream} provides techniques for tracking the progress of computation for each slice to reduce the overhead of reconstructing if information about the lineage graph is lost from memory. 

STream processing ELAsticity (Stela) is a system capable of optimising throughput after a scaling out/in operation and minimising the interruption to computation while the operation is being performed \cite{Xu:2016}. It uses Expected Throughput Percentage (ETP), which is a per-operator performance metric defined as the \textit{``final throughput that would be affected if the operator's processing speed were changed''}. While evaluation results demonstrate that ETP performs well as a post-scaling performance estimate, the work considers stateless operators whose migration can be performed without copying large amounts of application-related data. Stela is implemented as an extension to Storm's scheduler. Scale out/in operations are user-specified and are utilised to determine which operators are given more resources or which operators lose previously allocated resources.  

Hidalgo \etal \cite{Hidalgo:2016} employ operator fission to achieve elasticity by creating a processing graph that increases or decreases the number of processing operators to improve performance and resource utilisation. They introduce two algorithms to determine the state of an operator, namely a \textit{short-term} algorithm that evaluates load over short periods to detect traffic peaks; and (ii) a \textit{long-term} algorithm that finds traffic patterns. The short-term algorithm compares the actual load of an operator against upper and lower thresholds. The long-term algorithm uses a Markov chain based on operator history to evaluate state transitions over the analysed samples to define the matrix transition. The algorithm estimates for the next time-window the probability that an operator reaches one of the three possible states (\ie overloaded, underloaded, stable).

In the recent past, researchers and practitioners have also exploited the use of containers and lightweight resource virtualisation to perform migration of stream processing operators. Pahl and Lee \cite{Pahl:2015} review container technology as means to tackle elasticity in highly distributed environments comprising edge and cloud computing resources. Both containers and virtualisation technologies are useful when adjusting resource capacity during scale out/in operations, but containers are more lightweight, portable and provide more agility and flexibility when testing and deploying applications.

To support operator placement and migration in \ac{MCEP} systems, Ottenw\"alder \etal \cite{Ottenwalder:2013} present techniques that exploit system characteristics and predict mobility patterns for planning operator-state migration in advance. The envisioned infrastructure comprises a federation of distributed brokers whose hierarchy comprises a combination of cloud and fog resources. Mobile nodes connect to the nearest broker, and each operator along with its state are kept in their own virtual machine. The problem tackled consists of finding a sequence of placements and migrations for an application graph so that the network utilisation is minimised and the end-to-end latency requirements are met. The system performs an incremental placement where, a placement decision is enforced if its migration costs can be amortised by the gain of the next placement decision. A migration plan is dynamically updated for each operator and a \textit{time-graph model} is used for selecting migration targets and for negotiating the plans with dependent operators to find the minimum cost plans for each operator and reserve resources accordingly. The link load created by events is estimated considering the most recent traffic measurements, while latency is computed via regular ping messages or using Vivaldi coordinates \cite{Dabek:2004}.

\newcolumntype{C}[1]{>{\centering\let\newline\\\arraybackslash\hspace{0pt}}m{#1}@{\hspace{1mm}}}
\begin{table*}[!htb]
\centering 
\caption{Online techniques for elastic stream processing.}
\label{tab:online_elasticity} 
\footnotesize
  \begin{tabular}{C{22mm}C{15mm}C{20mm}C{35mm}C{15mm}C{35mm}}
    \toprule
    \multirow{3}{22mm}{\centering{\textbf{Solution}}} &
    \multirow{3}{15mm}{\centering{\textbf{Target Infrastructure}}} &
    \multirow{3}{20mm}{\centering{\textbf{Operator type}}} & 
    \multirow{3}{35mm}{\centering{\textbf{Metrics for Elasticity}}} & \\
    & & & & \multicolumn{2}{c}{\centering{\textbf{Elasticity}}}\\
    \cmidrule{5-6}
    & & & & \textbf{Type} & \textbf{Actions} \\
    \toprule
    Fernandez \etal \cite{FernandezScaleOut:2013} & cloud & stateful & Resource use (CPU) & horizontal & operator state check-pointing, fission \\ \midrule
    
    T-Storm \cite{XuTStorm:2014} & cluster & stateless & Resource use (CPU, inter-executor traffic load) & N/A & executor reassignment, topology rebalance \\ \midrule
    
    Adaptive Storm \cite{AnielloStormAdaptive:2013} & cluster & stateful bolts, stateless operators & Resource use (CPU, inter-node traffic) & N/A & executor placement, dynamic executor reassignment \\ \midrule
    
    Nephele SPE \cite{LohrmannLatency:2015} & cluster & stateless & System metrics (task and channel latency) & vertical & data batching, operator fission \\ \midrule
    
    \textsc{Esc} \cite{SatzgerESC:2011} & cloud & stateless${^1}$ & Resource use (machine load), system metrics (queue lengths) & horizontal & replace load balancing functions dynamically, operator fission \\ \midrule
    
    \acf{SC} \cite{GulisanoStreamCloud:2012} & cluster or private cloud${^2}$ & stateless and stateful & Resource use (CPU) & horizontal & query splitting and placement, compiler for query parallelisation \\ \midrule
    
    FUGU \cite{HeinzeFUGU:2014} & cloud & stateful & Resource use (CPU, network and memory consumption) & horizontal & operator migration, query placement \\ \midrule
    
     Gedik \etal \cite{GedikElastic:2014} & cluster & stateless and partitioned stateful & System metrics (congestion index, throughput) & vertical & operator fission, state check-pointing, operator migration  \\ \midrule
    
     \textit{ChronoStream} \cite{Wu:2015} & cloud & stateful & N/A & vertical and horizontal${^3}$ &  operator state check-pointing, replication, migration, parallelism  \\ \midrule
%
     Stela \cite{Xu:2016} & cloud & stateless & System metrics (impacted throughput) & horizontal${^3}$ & operator fission and migration \\ \midrule
    
    MigCEP \cite{Ottenwalder:2013} & cloud + fog & stateful & System metrics (load on event streams, inter-operator latency) & N/A & operator placement and migration \\ 
    \bottomrule
    \multicolumn{6}{l}{$^1$ \textsc{Esc} experiments consider only stateless operators.}\\
    \multicolumn{6}{l}{$^2$ Nodes must be pre-configured with \acl{SC}.}\\
    \multicolumn{6}{l}{$^3$ Execution of scale out/in operations are user-specified, not triggered by the system.}\\
    
  \end{tabular} 
\end{table*}

Table~\ref{tab:online_elasticity} summarises a select number of solutions that aim to provide elastic data stream processing. The table details the infrastructure targeted by the solutions (\ie cluster, cloud, fog); the types of operators considered (\ie stateless, stateful); the metrics monitored and taken into account when planning a scale out/in operation; the type of elasticity envisioned (\ie vertical or horizontal); and the elasticity actions performed during the execution of a scale out/in operation.

\subsection{Change and Burst Detection}
\label{subsec:change_detection}

Another approach that may be key to addressing elasticity in data stream processing is to use techniques to detect changes or bursts in the incoming data feeding a stream processing engine. This approach does not address elasticity per se, but it can be used with other techniques to trigger scale out/in operations such as adding or removing resources and employing graph adaptation.  
  
For instance, Zhu and Shasha \cite{ZhuBurstDetection:2003} introduce a shifted wavelet tree data structure for detecting bursts in aggregates of time series based data streams. They considered three types of sliding windows aggregates:

\begin{itemize}
\item \textit{Landmark windows:} aggregates are computed from a specific time point.
\item \textit{Sliding Windows:} aggregates are calculated based on a window of the last $n$ values.
\item \textit{Damped window:} the weights of data decrease exponentially into the past.
\end{itemize}

Krishnamurthy \etal \cite{KrishnamurthySketchChange:2003} propose a sketch data structure for summarising network traffic at multiple levels on top of which time series forecast models are applied to detect significant changes in flows that present large forecast errors. Previous work provides a literature review on the topic of change and burst deception. Tran \etal \cite{TranChangeDetection:2014}, for instance, present a survey on change detection techniques for data stream processing.

%% file: architectures.tex
\section{Distributed and Hybrid Architecture}
\label{sec:architectures}

Most distributed data stream processing systems have been traditionally designed for cluster environments. More recently, architectural models have emerged for more distributed environments spanning multiple data centres or for exploiting the edges of the Internet (\ie, edge and fog computing \cite{ChaoETSI:2015,SarkarAssessmentFog:2015}).Existing work aims to use the Internet edges by trying to place certain stream processing elements on micro data centres (often called Cloudlets \cite{SatyanarayananEdge:2017}) closer to where the data is generated \cite{Cardellini:2015}, transferring events to the cloud in batches \cite{TudoranJetStream:2016}, or by exploiting mobile devices in the fog for stream processing \cite{MoralesFogStream:2014}. Proposed architecture aims to place data analysis tasks at the edge of the Internet in order to reduce the amount of data transferred from sources to the cloud, improve the end-to-end latency, or offload certain analyses from the cloud \cite{ChanQuarksIBM:2016}. 

Despite many efforts on building infrastructure, such as adapting OpenStack to run on cloudlets, much of the existing work on stream processing, however, remains at a conceptual or architectural level without concrete software solutions or demonstrated scalability. Applications are still emerging. This section provides a non-exhaustive list of work regarding virtualisation infrastructure for stream processing, and placement and reconfiguration of stream processing applications.


\subsection{Lightweight Virtualisation and Containers}

Pahl \etal \cite{Pahl:2016} and Ismail \etal \cite{Ismail:2015} discussed the use of lightweight virtualisation and the need for orchestrating the deployment of containers as key requirements to address challenges in infrastructure comprising fog and cloud computing, such as improving application deployment speed, reducing overhead and data transferred over the network. Stream processing is often viewed as a motivating scenario. Yangui \etal \cite{Yangui:2016} propose a \ac{PaaS} architecture for cloud and fog integration. A proof-of-concept implementation is described, which extends Cloud Foundry \cite{CloudFoundry} to ease testing and deployment of applications whose components can be hosted either in the cloud or fog resources. 

Morabito and Beijar \cite{Morabito:2016} designed an \textit{Edge Computation Platform} for capillary networks \cite{Novo:2015} that takes advantage of lightweight containers to achieve resource elasticity. The solution exploits single board computers (\eg Raspberry Pi 2 B, and Odroid C1+) as gateways where certain functional blocks (\ie data compression and data processing) can be hosted. Similarly Petrolo \etal \cite{Petrolo:2016} focus on a gateway design for \ac{WSN} to optimise the communication and make use of the edges. The gateway, designed for a cloud of things, can manage semantic-like things and work as an end-point for data presentation to users.

Hochreiner \etal \cite{Hochreiner:2016} propose the VIenna ecosystem for elastic Stream Processing (VISP) which exploits the use of lightweight containers to enable application deployment on hybrid environments (\eg clouds and edge resources), a graphical interface for easy assemble of processing graphs, and reuse of processing operators. To achieve elasticity, the ecosystem runtime component monitors performance metrics of operators instances, the load on the message infrastructure, and introspection of the individual messages in the message queue.

\subsection{Application Placement and Reconfiguration}

Task scheduling considering hybrid scenarios has been investigated in other domains, such as mobile clouds \cite{GaiCloudlet:2016} and heterogeneous memory \cite{GaiMemory:2016}. For stream processing, Benoit \etal \cite{BenoitLinearChain:2013} show that scheduling linear chains of processing operators onto a cluster of heterogeneous hardware is an NP-Hard problem, whereas placement of virtual computing resources and network flows onto hybrid infrastructure has also been investigated in other contexts \cite{RohPlacement:2017}. 

For stream processing, Cardellini \etal \cite{Cardellini:2016} introduce an integer programming formulation that takes into account resource heterogeneity for the Optimal Distributed Stream Processing Problem (ODP). They propose an extension to Apache Storm to incorporate an ODP-based scheduler, which estimates networks latency via a network coordination system built using the Vivaldi algorithm \cite{Dabek:2004}. It has been shown, however, that assigning stream processing operators on \acp{VM} and placing them across multiple geographically distributed data centres while minimising the overall inter data-centre communication cost, can often be classified as an NP-Hard \cite{GuCostStream:2016} problem or even NP-Complete \cite{Tziritas:2016}. Over time, however, cost-aware heuristics have been proposed for assigning stream processing operators to \acp{VM} placed across multiple data centres \cite{GuCostStream:2016,ChenStreamCost:2016}.

Sajjad and Danniswara \cite{Sajjad:2016} introduce a stream processing solution, \ie SpanEdge, that uses central and edge data centres. SpanEdge follows a master-worker architecture with \textit{hub} and \textit{spoke} workers, where a \textit{hub-worker} is hosted at a central data centre and a \textit{spoke-worker} at an edge data centre. SpanEdge also enables global and local tasks, and its scheduler attempts to place local tasks near the edges and global tasks at central data centres to minimise the impact of the latency of \ac{WAN} links interconnecting the data centres.

Mehdipour \etal \cite{Mehdipour:2016} introduce a hierarchical architecture for processing streamlined data using fog and cloud resources. They focus on minimising communication requirements between fog and cloud when processing data from \acp{IoT} devices. Shen \etal \cite{Shen:2015} advocate the use of Cisco's Connected Streaming Analytics (CSA) for conceiving an architecture for handling data stream processing queries for IoT applications by exploiting data centre and edge computing resources. \ac{CSA} provides a query language for continuous queries over streams. 

Geelytics is a system tailored for \ac{IoT} environments that comprise multiple geographically distributed data producers, result consumers, and computing resources that can be hosted either on the cloud or at the network edges \cite{ChengEdgeStream:2016}. Geelytics follows a master-worker architecture with a publish/subscribe service. Similarly to other data stream processing systems, Geelytics structures applications as \acp{DAG} of operators. Unlike other systems, however, it enables scoped tasks, where a user specifies the scope granularity of each task comprising the processing graph. The scope granularity of tasks and data-consumer scoped subscriptions are used to devise the execution plan and deploy the resulting tasks according to the geographical location of data producers.

%% file: gap_analysis.tex
\section{Future Directions}
\label{sec:future_directions}

Organisations often demand not only online processing of large amounts of streaming data, but also solutions that can perform computations on large data sets by using models such as MapReduce. As a result, big data processing solutions employed by large organisations exploit hybrid execution models (\eg using batch and online execution) that can span multiple data centres. In addition to providing elasticity for computing and storage resources, ideally, a big data processing service should be able to allocate and release resources on demand. This section highlights some future directions.

\subsection{SDN and In-Transit Processing}

Networks are becoming increasingly programmable by using several solutions such as \ac{SDN} \cite{KreutzSDNSurvey:2015} and \ac{NFV}, which can provide mechanisms required for allocating network capacity for certain data flows both within and across data centres with certain computing operations been performed in-network. In-transit stream processing can be carried out where certain processing elements, or operators, are placed along the network interconnecting data sources and the cloud. This approach raises security and resource management challenges. In scenarios such as \ac{IoT}, having components that perform processing along the path from data sources to the cloud can increase the number of hops susceptible to attacks. Managing task scheduling and allocation of heterogeneous resources whilst offering the elasticity with which cloud users are accustomed is also difficult as adapting an application to current resource and network conditions may require migrating elements of a data flow that often maintain state.

Most of the existing work on multi-operator placement considered network metrics such as latency and bandwidth while proposing decentralised algorithms, without taking into account that the network can be programmed and capacity allocated to certain network flows. The interplay between hybrid models and \ac{SDN} as well as joint optimisation of application placement and flow routing can be better explored. The optimal placement of data processing elements and adaptation of data flow graphs, however, are hard problems. 

In addition to placing operators on heterogeneous environments, a key issue is deciding which operators are worth placing on edge computing resources and which should remain in the cloud. Emerging cognitive assistance scenarios \cite{HaWearable:2014} offer interesting use cases where machine learning models can be trained on the cloud, and once trained they can be deployed on edge computing resources. The challenge, however, is to identify eventual concept drifts that in turn require retraining a model and potentially adapting the execution data flow.   

\subsection{Programming Models for Hybrid and\\ Highly Distributed Architecture}

Frameworks that provide high-level programming abstractions have been introduced in recent past to ease the development and deployment of big data applications that use hybrid models \cite{BoykinSummingbird:2014,GoogleDataFlow}. Platform bindings have been provided to deploy applications developed using these abstractions on the infrastructure provided by commercial public cloud providers and open source solutions. Although such solutions are often restricted to a single cluster or data centre, efforts have been made to leverage resources from the edges of the Internet to perform distributed queries \cite{VulimiriWANalyticsNSDI:2015} or to push frequently-performed analytics tasks to edge resources \cite{ChengEdgeStream:2016}. With the growing number of scenarios where data is collected by a plethora of devices, such as in \ac{IoT} and smart cities, and requires processing under low latency, solutions are increasingly exploiting resources available at the edges of the Internet (\ie edge and fog computing). In addition to providing means to place data processing tasks in such environments while minimising the use of network resources and latency, efficient methods to manage resource elasticity in these scenarios should be investigated. Moreover, high-level programming abstractions and bindings to platforms capable of deploying and managing resources under such highly distributed scenarios are desirable.

Under the Apache Beam project \cite{ApacheBeam}, efforts have been made towards providing a unified SDK while enabling processing pipelines to be executed on distributed processing back-ends such as Apache Spark \cite{ZahariaRDD:2012} and Apache Flink \cite{ApacheFlink}. Beam is particularly useful for embarrassingly parallel applications. There is still a lack of unified SDKs that simplify application development covering the whole spectrum, from data collection at the internet edges to processing at micro data centres (more closely located to the Internet edges) and data centres. Concerning resource management for such environments, several challenges arise regarding the network infrastructure and resource heterogeneity. Despite the challenges regarding state management for stream processing systems, container-based solutions could facilitate the deployment and elasticity management under such environments \cite{Kubernets}, and solutions such as Apache Quarks/Edgent \cite{ApacheEdgent} can be leveraged to perform certain analyses at the Internet edges.

%% file: survey.bbl
\begin{thebibliography}{100}
\expandafter\ifx\csname url\endcsname\relax
  \def\url#1{\texttt{#1}}\fi
\expandafter\ifx\csname urlprefix\endcsname\relax\def\urlprefix{URL }\fi
\expandafter\ifx\csname href\endcsname\relax
  \def\href#1#2{#2} \def\path#1{#1}\fi

\bibitem{CiscoMSE:2012}
\href{http://www.cisco.com/en/US/prod/collateral/wireless/c36_696714_00_copenhagen_airport_cs.pdf}{{Unlocking
  Game-Changing Wireless Capabilities: Cisco and {SITA} help {Copenhagen
  Airport} Develop New Services for Transforming the Passenger Experience}},
  Customer case study, CISCO (2012).
\newline\urlprefix\url{http://www.cisco.com/en/US/prod/collateral/wireless/c36_696714_00_copenhagen_airport_cs.pdf}

\bibitem{attentionshoppers:2013}
S.~Clifford, Q.~Hardy,
  \href{http://www.nytimes.com/2013/07/15/business/attention-shopper-stores-are-tracking-your-cell.html}{Attention,
  shoppers: Store is tracking your cell}, New York Times (2013).
\newline\urlprefix\url{http://www.nytimes.com/2013/07/15/business/attention-shopper-stores-are-tracking-your-cell.html}

\bibitem{HaWearable:2014}
K.~Ha, Z.~Chen, W.~Hu, W.~Richter, P.~Pillai, M.~Satyanarayanan,
  \href{http://doi.acm.org/10.1145/2594368.2594383}{Towards wearable cognitive
  assistance}, in: 12th Annual International Conference on Mobile Systems,
  Applications, and Services, MobiSys '14, ACM, New York, USA, 2014, pp.
  68--81.
\newblock \href {http://dx.doi.org/10.1145/2594368.2594383}
  {\path{doi:10.1145/2594368.2594383}}.
\newline\urlprefix\url{http://doi.acm.org/10.1145/2594368.2594383}

\bibitem{AssuncaoJPDC:2015}
M.~D. de~Assuncao, R.~N. Calheiros, S.~Bianchi, M.~A.~S. Netto, R.~Buyya, Big
  data computing and clouds: Trends and future directions, Journal of Parallel
  and Distributed Computing 79--80~(0) (2015) 3--15.

\bibitem{AtzoriIoTSurvey:2010}
L.~Atzori, A.~Iera, G.~Morabito, The internet of things: A survey, Computer
  Networks 54~(15) (2010) 2787--2805.

\bibitem{RettigOnline:2015}
L.~Rettig, M.~Khayati, P.~Cudr{\'e}-Mauroux, M.~Pi{\'o}rkowski, Online anomaly
  detection over big data streams, in: IEEE International Conference on Big
  Data ({Big Data 2015}), IEEE, Santa Clara, USA, 2015, pp. 1113--1122.

\bibitem{ArmbrustCloud:2009}
M.~Armbrust, A.~Fox, R.~Griffith, A.~D. Joseph, R.~H. Katz, A.~Konwinski,
  G.~Lee, D.~A. Patterson, A.~Rabkin, I.~Stoica, M.~Zaharia, {Above the Clouds:
  A {Berkeley} View of {C}loud Computing}, Technical report UCB/EECS-2009-28,
  Electrical Engineering and Computer Sciences, University of California at
  Berkeley, Berkeley, USA (February 2009).

\bibitem{BoykinSummingbird:2014}
O.~Boykin, S.~Ritchie, I.~O'Connell, J.~Lin, Summingbird: A framework for
  integrating batch and online {MapReduce} computations, Proceedings of the
  {VLDB} Endowment 7~(13) (2014) 1441--1451.

\bibitem{GoogleDataFlow}
{Google Cloud Dataflow}, https://cloud.google.com/dataflow/ (2015).

\bibitem{ChaoETSI:2015}
Y.~C. Hu, M.~Patel, D.~Sabella, N.~Sprecher, V.~Young, Mobile edge computing: A
  key technology towards {5G}, Whitepaper ETSI White Paper No. 11, European
  Telecommunications Standards Institute ({ETSI}) (September 2015).

\bibitem{HuEdge:2016}
W.~Hu, Y.~Gao, K.~Ha, J.~Wang, B.~Amos, Z.~Chen, P.~Pillai, M.~Satyanarayanan,
  \href{http://doi.acm.org/10.1145/2967360.2967369}{Quantifying the impact of
  edge computing on mobile applications}, in: 7th {ACM SIGOPS} Asia-Pacific
  Workshop on Systems, APSys '16, ACM, New York, USA, 2016, pp. 5:1--5:8.
\newblock \href {http://dx.doi.org/10.1145/2967360.2967369}
  {\path{doi:10.1145/2967360.2967369}}.
\newline\urlprefix\url{http://doi.acm.org/10.1145/2967360.2967369}

\bibitem{ApacheEdgent}
{Apache Edgent}, https://edgent.apache.org (2017).

\bibitem{PisaniSBAC:2017}
F.~Pisani, J.~R. Brunetta, V.~M. do~Rosario, E.~Borin, Beyond the fog: Bringing
  cross-platform code execution to constrained iot devices, in: 29th
  International Symposium on Computer Architecture and High Performance
  Computing ({SBAC-PAD 2017}), Campinas, Brazil, 2017, pp. 17--24.

\bibitem{ZhaoSurveyStreams:2017}
X.~Zhao, S.~Garg, C.~Queiroz, R.~Buyya, Software Architecture for Big Data and
  the Cloud, Elsevier -- Morgan Kaufmann, 2017, Ch. A Taxonomy and Survey of
  Stream Processing Systems.

\bibitem{EllisRealTimeStreaming:2014}
B.~Ellis, Real-time analytics: Techniques to Analyze and Visualize Streaming
  Data, John Wiley \& Sons, Indianapolis, USA, 2014.

\bibitem{AllenStormApplied:2015}
S.~T. Allen, M.~Jankowski, P.~Pathirana, Storm Applied: Strategies for
  Real-time Event Processing, 1st Edition, Manning Publications Co., Greenwich,
  USA, 2015.

\bibitem{LiuStreamIoT:2016}
X.~Liu, A.~V. Dastjerdi, R.~Buyya, Internet of Things: Principles and
  Paradigms, Morgan Kaufmann, Burlington, USA, 2016, Ch. Stream Processing in
  {IoT}: Foundations, State-of-the-art, and Future Directions.

\bibitem{CentenaroIoT:2016}
M.~Centenaro, L.~Vangelista, A.~Zanella, M.~Zorzi, Long-range communications in
  unlicensed bands: the rising stars in the iot and smart city scenarios,
  Vol.~23, 2016, pp. 60--67.
\newblock \href {http://dx.doi.org/10.1109/MWC.2016.7721743}
  {\path{doi:10.1109/MWC.2016.7721743}}.

\bibitem{ApacheThrift}
{Apache Thrift}, https://thrift.apache.org/ (2016).

\bibitem{ProtoBuff}
{Protocol Buffers}, https://developers.google.com/protocol-buffers/ (2016).

\bibitem{ApacheActiveMQ}
{Apache ActiveMQ}, http://activemq.apache.org/ (2016).

\bibitem{RabiitMQ}
{RabbitMQ}, https://www.rabbitmq.com/ (2016).

\bibitem{Kestrel}
Kestrel, https://github.com/twitter-archive/kestrel (2016).

\bibitem{ApacheKafka}
{Apache Kafka}, http://kafka.apache.org/ (2016).

\bibitem{DistributedLog}
{DistributedLog}, http://distributedlog.io/ (2016).

\bibitem{AmazonFirehose}
{Amazon Kinesis Firehose}, https://aws.amazon.com/kinesis/firehose/ (2015).

\bibitem{AzureIoTHub}
{Azure IoT Hub}, https://azure.microsoft.com/en-us/services/iot-hub/ (2016).

\bibitem{JeffreyMapReduce:2008}
J.~Dean, S.~Ghemawat, {MapReduce}: Simplified data processing on large
  clusters, Communications of the ACM 51~(1).

\bibitem{BorthakurFacebookHadoop:2011}
D.~Borthakur, J.~Gray, J.~S. Sarma, K.~Muthukkaruppan, N.~Spiegelberg,
  H.~Kuang, K.~Ranganathan, D.~Molkov, A.~Menon, S.~Rash, R.~Schmidt, A.~Aiyer,
  {Apache Hadoop Goes Realtime at {Facebook}}, in: Proceedings of the {ACM
  SIGMOD} International Conference on Management of Data ({SIGMOD 2011}), ACM,
  New York, USA, 2011, pp. 1071--1080.

\bibitem{ChenEnergyHadoop:2012}
Y.~Chen, S.~Alspaugh, D.~Borthakur, R.~Katz, Energy efficiency for large-scale
  {MapReduce} workloads with significant interactive analysis, in: 7th {ACM}
  European Conference on Computer Systems ({EuroSys 2012}), ACM, New York, USA,
  2012, pp. 43--56.

\bibitem{HanNOSQLSurvey:2011}
J.~Han, H.~E, G.~Le, J.~Du, Survey on {NoSQL} database, in: 6th International
  Conference on Pervasive Computing and Applications ({ICPCA 2011}), IEEE, Port
  Elizabeth, South Africa, 2011, pp. 363--366.

\bibitem{MuthukrishnanStreams:2005}
S.~Muthukrishnan, Data streams: Algorithms and applications, Now Publishers
  Inc., 2005.

\bibitem{Babcock:2002}
B.~Babcock, S.~Babu, M.~Datar, R.~Motwani, J.~Widom,
  \href{http://doi.acm.org/10.1145/543613.543615}{Models and issues in data
  stream systems}, in: 21st {ACM SIGMOD-SIGACT-SIGART} Symposium on Principles
  of Database Systems, PODS '02, ACM, New York, USA, 2002, pp. 1--16.
\newblock \href {http://dx.doi.org/10.1145/543613.543615}
  {\path{doi:10.1145/543613.543615}}.
\newline\urlprefix\url{http://doi.acm.org/10.1145/543613.543615}

\bibitem{GolabIssuesStream:2003}
L.~Golab, M.~T. \"{O}zsu,
  \href{http://doi.acm.org/10.1145/776985.776986}{Issues in data stream
  management}, SIGMOD Record 32~(2) (2003) 5--14.
\newblock \href {http://dx.doi.org/10.1145/776985.776986}
  {\path{doi:10.1145/776985.776986}}.
\newline\urlprefix\url{http://doi.acm.org/10.1145/776985.776986}

\bibitem{KulkarniHeron:2015}
S.~Kulkarni, N.~Bhagat, M.~Fu, V.~Kedigehalli, C.~Kellogg, S.~Mittal, J.~M.
  Patel, K.~Ramasamy, S.~Taneja, {Twitter Heron}: Stream processing at scale,
  in: {ACM SIGMOD} International Conference on Management of Data, SIGMOD '15,
  ACM, New York, USA, 2015, pp. 239--250.

\bibitem{GedikPipelinedFission:2016}
B.~Gedik, H.~\"{O}zsema, O.~\"{O}zt\"{u}rk, Pipelined fission for stream
  programs with dynamic selectivity and partitioned state, Journal of Parallel
  and Distributed Computing 96 (2016) 106--120.
\newblock \href
  {http://dx.doi.org/http://dx.doi.org/10.1016/j.jpdc.2016.05.003}
  {\path{doi:http://dx.doi.org/10.1016/j.jpdc.2016.05.003}}.

\bibitem{TangAutoPipelining:2013}
Y.~Tang, B.~Gedik, Autopipelining for data stream processing, IEEE Transactions
  on Parallel and Distributed Systems 24~(12) (2013) 2344--2354.
\newblock \href {http://dx.doi.org/10.1109/TPDS.2012.333}
  {\path{doi:10.1109/TPDS.2012.333}}.

\bibitem{SattlerElastic:2013}
K.-U. Sattler, F.~Beier, Towards elastic stream processing: Patterns and
  infrastructure, in: 1st International Workshop on Big Dynamic Distributed
  Data ({BD3}), Riva del Garda, Italy, 2013, pp. 49--54.

\bibitem{WuCEPLanguage:2006}
E.~Wu, Y.~Diao, S.~Rizvi, High-performance complex event processing over
  streams, in: {ACM SIGMOD} International Conference on Management of Data,
  SIGMOD '06, ACM, New York, USA, 2006, pp. 407--418.

\bibitem{GyllstromSASE:2007}
D.~Gyllstrom, E.~Wu, H.~Chae, Y.~Diao, P.~Stahlberg, G.~Anderson, {SASE:}
  complex event processing over streams (demo), in: Third Biennial Conference
  on Innovative Data Systems Research ({CIDR} 2007), 2007, pp. 407--411.

\bibitem{HeComet:2010}
B.~He, M.~Yang, Z.~Guo, R.~Chen, B.~Su, W.~Lin, L.~Zhou, Comet: Batched stream
  processing for data intensive distributed computing, in: 1st ACM Symposium on
  Cloud Computing, SoCC '10, ACM, New York, USA, 2010, pp. 63--74.
\newblock \href {http://dx.doi.org/10.1145/1807128.1807139}
  {\path{doi:10.1145/1807128.1807139}}.

\bibitem{Sajjad:2016}
H.~P. Sajjad, K.~Danniswara, A.~Al-Shishtawy, V.~Vlassov, {SpanEdge}: Towards
  unifying stream processing over central and near-the-edge data centers, in:
  {IEEE/ACM} Symposium on Edge Computing ({SEC}), 2016, pp. 168--178.

\bibitem{ChanQuarksIBM:2016}
S.~Chan,
  \href{https://www.ibm.com/blogs/bluemix/2016/06/better-analytics-with-apache-quarks/}{Apache
  quarks, watson, and streaming analytics: Saving the world, one smart
  sprinkler at a time}, Bluemix Blog (June 2016).
\newline\urlprefix\url{https://www.ibm.com/blogs/bluemix/2016/06/better-analytics-with-apache-quarks/}

\bibitem{HirzelSPL:2017}
M.~Hirzel, S.~Schneider, B.~Gedik,
  \href{http://doi.acm.org/10.1145/3039207}{{SPL}: An extensible language for
  distributed stream processing}, ACM Trans. Program. Lang. Syst. 39~(1) (2017)
  5:1--5:39.
\newblock \href {http://dx.doi.org/10.1145/3039207}
  {\path{doi:10.1145/3039207}}.
\newline\urlprefix\url{http://doi.acm.org/10.1145/3039207}

\bibitem{NettoScaling:2014}
M.~A.~S. Netto, C.~Cardonha, R.~Cunha, M.~D. de~Assuncao, Evaluating
  auto-scaling strategies for cloud computing environments, in: 22nd {IEEE}
  Int. Symp. on Modeling, Analysis and Simulation of Computer and
  Telecommunication Systems ({MASCOTS 2014}), IEEE, 2014, pp. 187--196.

\bibitem{TolosanaMultiTenancyStreas:2016}
R.~Tolosana-Calasanz, J.~çngel Ba–ares, C.~Pham, O.~F. Rana, Resource
  management for bursty streams on multi-tenancy cloud environments, Future
  Generation Computer Systems 55 (2016) 444--459.
\newblock \href
  {http://dx.doi.org/https://doi.org/10.1016/j.future.2015.03.012}
  {\path{doi:https://doi.org/10.1016/j.future.2015.03.012}}.

\bibitem{ChenNiagaraCQ:2000}
J.~Chen, D.~J. DeWitt, F.~Tian, Y.~Wang, {NiagaraCQ}: A scalable continuous
  query system for internet databases, in: {ACM SIGMOD} International
  Conference on Management of Data, SIGMOD '00, ACM, New York, USA, 2000, pp.
  379--390.

\bibitem{ArasuSTREAM:2004}
A.~Arasu, B.~Babcock, S.~Babu, J.~Cieslewicz, M.~Datar, K.~Ito, R.~Motwani,
  U.~Srivastava, J.~Widom, Stream: The stanford data stream management system,
  Book chapter, Stanford InfoLab (2004).

\bibitem{BabcockChain:2003}
B.~Babcock, S.~Babu, R.~Motwani, M.~Datar, Chain: Operator scheduling for
  memory minimization in data stream systems, in: {ACM SIGMOD} International
  Conference on Management of Data, SIGMOD '03, ACM, New York, USA, 2003, pp.
  253--264.

\bibitem{AbadiAurora:2003}
D.~J. Abadi, D.~Carney, U.~\c{C}etintemel, M.~Cherniack, C.~Convey, S.~Lee,
  M.~Stonebraker, N.~Tatbul, S.~Zdonik, Aurora: A new model and architecture
  for data stream management, Vol.~12, Springer-Verlag New York, Inc.,
  Secaucus, USA, 2003, pp. 120--139.

\bibitem{BalazinskaMedusa:2005}
M.~Balazinska, H.~Balakrishnan, M.~Stonebraker, Contract-based load management
  in federated distributed systems, in: 1st Symposium on Networked Systems
  Design and Implementation ({NSDI}), USENIX Association, San Francisco, USA,
  2004, pp. 197--210.

\bibitem{TatbulFIT:2007}
N.~Tatbul, U.~\c{C}etintemel, S.~Zdonik, Staying {FIT}: Efficient load shedding
  techniques for distributed stream processing, in: 33rd International
  Conference on Very Large Data Bases, VLDB '07, VLDB Endowment, 2007, pp.
  159--170.

\bibitem{AbadiBorealis:2005}
D.~J. Abadi, Y.~Ahmad, M.~Balazinska, U.~Cetintemel, M.~Cherniack, J.-H. Hwang,
  W.~Lindner, A.~Maskey, A.~Rasin, E.~Ryvkina, N.~Tatbul, Y.~Xing, S.~Zdonik,
  The design of the borealis stream processing engine, in: Conference on
  Innovative Data Systems Research (CIDR), Vol.~5, 2005, pp. 277--289.

\bibitem{ZooKeeper}
{Apache Zookeeper}, http://zookeeper.apache.org/ (2016).

\bibitem{BasantaRealTime:2015}
P.~Basanta-Val, N.~Fern‡ndez-Garc'a, A.~Wellings, N.~Audsley, Improving the
  predictability of distributed stream processors, Future Generation Computer
  Systems 52 (2015) 22--36, special Section: Cloud Computing: Security, Privacy
  and Practice.
\newblock \href
  {http://dx.doi.org/http://dx.doi.org/10.1016/j.future.2015.03.023}
  {\path{doi:http://dx.doi.org/10.1016/j.future.2015.03.023}}.

\bibitem{ToshniwalStormAtTwitter:2014}
A.~Toshniwal, S.~Taneja, A.~Shukla, K.~Ramasamy, J.~M. Patel, S.~Kulkarni,
  J.~Jackson, K.~Gade, M.~Fu, J.~Donham, N.~Bhagat, S.~Mittal, D.~Ryaboy,
  Storm@twitter, in: {ACM SIGMOD} International Conference on Management of
  Data, SIGMOD '14, ACM, New York, USA, 2014, pp. 147--156.

\bibitem{HindmanMesos:2011}
B.~Hindman, A.~Konwinski, M.~Zaharia, A.~Ghodsi, A.~D. Joseph, R.~Katz,
  S.~Shenker, I.~Stoica, {Mesos:} a platform for fine-grained resource sharing
  in the data center., in: NSDI, Vol.~11, 2011, pp. 22--22.

\bibitem{AmazonECS}
{Amazon EC2 Container Service}, https://aws.amazon.com/ecs/ (2015).

\bibitem{NeumeyerS4:2010}
L.~Neumeyer, B.~Robbins, A.~Nair, A.~Kesari, {S4:} distributed stream computing
  platform, in: {IEEE} International Conference on Data Mining Workshops
  ({ICDMW}), 2010, pp. 170--177.

\bibitem{ApacheSamza}
{Apache Samza}, https://samza.apache.org (2017).

\bibitem{VavilapalliYARN:2013}
V.~K. Vavilapalli, A.~C. Murthy, C.~Douglas, S.~Agarwal, M.~Konar, R.~Evans,
  T.~Graves, J.~Lowe, H.~Shah, S.~Seth, B.~Saha, C.~Curino, O.~O'Malley,
  S.~Radia, B.~Reed, E.~Baldeschwieler, {Apache Hadoop YARN}: Yet another
  resource negotiator, in: 4th Annual Symposium on Cloud Computing, SOCC '13,
  ACM, New York, USA, 2013, pp. 5:1--5:16.
\newblock \href {http://dx.doi.org/10.1145/2523616.2523633}
  {\path{doi:10.1145/2523616.2523633}}.

\bibitem{ApacheFlink}
{Apache Flink}, http://flink.apache.org/ (2015).

\bibitem{FlinkGelly}
\href{https://ci.apache.org/projects/flink/flink-docs-release-1.3/dev/libs/gelly/iterative_graph_processing.html}{Apache
  flink -- iterative graph processing}, API Documentation (2017).
\newline\urlprefix\url{https://ci.apache.org/projects/flink/flink-docs-release-1.3/dev/libs/gelly/iterative_graph_processing.html}

\bibitem{ZahariaRDD:2012}
M.~Zaharia, M.~Chowdhury, T.~Das, A.~Dave, J.~Ma, M.~McCauley, M.~J. Franklin,
  S.~Shenker, I.~Stoica, Resilient distributed datasets: A fault-tolerant
  abstraction for in-memory cluster computing, in: 9th {USENIX} Conference on
  Networked Systems Design and Implementation, NSDI'12, USENIX Association,
  Berkeley, USA, 2012, pp. 2--2.

\bibitem{ZahariaDiscretizedStreams:2013}
M.~Zaharia, T.~Das, H.~Li, T.~Hunter, S.~Shenker, I.~Stoica, Discretized
  streams: Fault-tolerant streaming computation at scale, in: 24th {ACM}
  Symposium on Operating Systems Principles, SOSP '13, ACM, New York, USA,
  2013, pp. 423--438.

\bibitem{ShaFlux:2003}
M.~A. Shah, J.~M. Hellerstein, S.~Chandrasekaran, M.~J. Franklin, Flux: An
  adaptive partitioning operator for continuous query systems, in: 19th
  International Conference on Data Engineering ({ICDE 2003}), {IEEE} Computer
  Society, 2003, pp. 25--36.

\bibitem{Wu:2015}
Y.~Wu, K.~L. Tan, {ChronoStream}: Elastic stateful stream computation in the
  cloud, in: 2015 IEEE 31st International Conference on Data Engineering, 2015,
  pp. 723--734.

\bibitem{AminiSPC:2006}
L.~Amini, H.~Andrade, R.~Bhagwan, F.~Eskesen, R.~King, P.~Selo, Y.~Park,
  C.~Venkatramani, {SPC}: A distributed, scalable platform for data mining, in:
  4th International Workshop on Data Mining Standards, Services and Platforms,
  DMSSP '06, ACM, New York, USA, 2006, pp. 27--37.

\bibitem{SatzgerESC:2011}
B.~Satzger, W.~Hummer, P.~Leitner, S.~Dustdar, Esc: Towards an elastic stream
  computing platform for the cloud, in: {IEEE} International Conference on
  Cloud Computing ({CLOUD}), 2011, pp. 348--355.

\bibitem{QianTimeStream:2013}
Z.~Qian, Y.~He, C.~Su, Z.~Wu, H.~Zhu, T.~Zhang, L.~Zhou, Y.~Yu, Z.~Zhang,
  Timestream: Reliable stream computation in the cloud, in: 8th ACM European
  Conference on Computer Systems, EuroSys '13, ACM, New York, USA, 2013, pp.
  1--14.
\newblock \href {http://dx.doi.org/10.1145/2465351.2465353}
  {\path{doi:10.1145/2465351.2465353}}.

\bibitem{SahaTez:2015}
B.~Saha, H.~Shah, S.~Seth, G.~Vijayaraghavan, A.~Murthy, C.~Curino, {Apache
  Tez}: A unifying framework for modeling and building data processing
  applications, in: 2015 ACM SIGMOD International Conference on Management of
  Data, SIGMOD '15, ACM, New York, USA, 2015, pp. 1357--1369.
\newblock \href {http://dx.doi.org/10.1145/2723372.2742790}
  {\path{doi:10.1145/2723372.2742790}}.

\bibitem{AkidauMillWheel:2013}
T.~Akidau, A.~Balikov, K.~Bekiro\u{g}lu, S.~Chernyak, J.~Haberman, R.~Lax,
  S.~McVeety, D.~Mills, P.~Nordstrom, S.~Whittle, Millwheel: Fault-tolerant
  stream processing at internet scale, Vol.~6, VLDB Endowment, 2013, pp.
  1033--1044.
\newblock \href {http://dx.doi.org/10.14778/2536222.2536229}
  {\path{doi:10.14778/2536222.2536229}}.

\bibitem{HuELF:2014}
L.~Hu, K.~Schwan, H.~Amur, X.~Chen, {ELF}: Efficient lightweight fast stream
  processing at scale., in: {USENIX} Annual Technical Conference, {USENIX}
  Association, Philadelphia, USA, 2014, pp. 25--36.

\bibitem{CloudWatch}
{Amazon CloudWatch}, https://aws.amazon.com/cloudwatch/ (2015).

\bibitem{ApacheBeam}
{Apache Beam}, http://beam.incubator.apache.org/ (2016).

\bibitem{GoogleComputeEngine}
{Google Compute Engine}, https://cloud.google.com/compute/ (2015).

\bibitem{GoogleCloudStorage}
{Google Cloud Storage}, https://cloud.google.com/storage/ (2015).

\bibitem{AzureStreamAnalytics}
{Azure Stream Analytics},
  https://azure.microsoft.com/en-us/services/stream-analytics/ (2015).

\bibitem{LoridoBotranElasticity:2014}
T.~Lorido-Botran, J.~Miguel-Alonso, J.~A. Lozano, A review of auto-scaling
  techniques for elastic applications in cloud environments, Journal of Grid
  Computing 12~(4) (2014) 559--592.

\bibitem{HirzelStreamOptimisation:2014}
M.~Hirzel, R.~Soul{\'e}, S.~Schneider, B.~Gedik, R.~Grimm, A catalog of stream
  processing optimizations, ACM Computing Surveys 46~(4) (2014) 1--34.

\bibitem{LakshmananPlaceStrategies:2008}
G.~T. Lakshmanan, Y.~Li, R.~Strom, Placement strategies for internet-scale data
  stream systems, {IEEE} Internet Computing 12~(6) (2008) 50--60.

\bibitem{PengRStorm:2015}
B.~Peng, M.~Hosseini, Z.~Hong, R.~Farivar, R.~Campbell, R-storm: Resource-aware
  scheduling in storm, in: 16th Annual Middleware Conference, Middleware '15,
  ACM, New York, USA, 2015, pp. 149--161.

\bibitem{PietzuchOpPlaceNetwork:2006}
P.~Pietzuch, J.~Ledlie, J.~Shneidman, M.~Roussopoulos, M.~Welsh, M.~Seltzer,
  Network-aware operator placement for stream-processing systems, in: 22nd
  International Conference on Data Engineering ({ICDE'06}), 2006, pp. 49--49.

\bibitem{ZhouOpPlacement:2006}
Y.~Zhou, B.~C. Ooi, K.-L. Tan, J.~Wu, Efficient Dynamic Operator Placement in a
  Locally Distributed Continuous Query System, Springer Berlin Heidelberg,
  Berlin, Heidelberg, 2006, pp. 54--71.

\bibitem{AhmadOpPlacement:2004}
Y.~Ahmad, U.~\c{C}etintemel, Network-aware query processing for stream-based
  applications, in: 13th International Conference on Very Large Data Bases --
  Volume 30, VLDB '04, VLDB Endowment, 2004, pp. 456--467.

\bibitem{FernandezScaleOut:2013}
R.~C. Fernandez, M.~Migliavacca, E.~Kalyvianaki, P.~Pietzuch, Integrating scale
  out and fault tolerance in stream processing using operator state management,
  in: {ACM SIGMOD} International Conference on Management of Data, SIGMOD '13,
  ACM, New York, USA, 2013, pp. 725--736.

\bibitem{HeinzeFUGU:2014}
T.~Heinze, Z.~Jerzak, G.~Hackenbroich, C.~Fetzer, Latency-aware elastic scaling
  for distributed data stream processing systems, in: 8th {ACM} International
  Conference on Distributed Event-Based Systems, DEBS '14, ACM, New York, USA,
  2014, pp. 13--22.

\bibitem{ViglasRateOptimisation:2002}
S.~D. Viglas, J.~F. Naughton, Rate-based query optimization for streaming
  information sources, in: {ACM SIGMOD} International Conference on Management
  of Data, SIGMOD '02, ACM, New York, USA, 2002, pp. 37--48.

\bibitem{LohrmannLatency:2015}
B.~Lohrmann, P.~Janacik, O.~Kao, Elastic stream processing with latency
  guarantees, in: 35th {IEEE} International Conference on Distributed Computing
  Systems ({ICDCS}), 2015, pp. 399--410.

\bibitem{KrishnamurthySketchChange:2003}
B.~Krishnamurthy, S.~Sen, Y.~Zhang, Y.~Chen, Sketch-based change detection:
  Methods, evaluation, and applications, in: 3rd {ACM SIGCOMM} Conference on
  Internet Measurement, IMC '03, ACM, New York, USA, 2003, pp. 234--247.

\bibitem{XuTStorm:2014}
J.~Xu, Z.~Chen, J.~Tang, S.~Su, {T-Storm}: Traffic-aware online scheduling in
  storm, in: {IEEE} 34th International Conference on Distributed Computing
  Systems ({ICDCS}), 2014, pp. 535--544.

\bibitem{AnielloStormAdaptive:2013}
L.~Aniello, R.~Baldoni, L.~Querzoni, Adaptive online scheduling in storm (2013)
  207--218.

\bibitem{GulisanoStreamCloud:2012}
V.~Gulisano, R.~Jim\'{e}nez-Peris, M.~{Pati\~{n}o-Mart\'{i}nez}, C.~Soriente,
  P.~Valduriez, {StreamCloud}: An elastic and scalable data streaming system,
  {IEEE} Transactions on Parallel and Distributed Systems 23~(12) (2012)
  2351--2365.

\bibitem{GedikElastic:2014}
B.~Gedik, S.~Schneider, M.~Hirzel, K.-L. Wu, Elastic scaling for data stream
  processing, {IEEE} Transactions on Parallel and Distributed Systems 25~(6)
  (2014) 1447--1463.

\bibitem{Xu:2016}
L.~Xu, B.~Peng, I.~Gupta, Stela: Enabling stream processing systems to scale-in
  and scale-out on-demand, IEEE International Conference on Cloud Engineering
  ({IC2E 2016}) 00 (2016) 22--31.

\bibitem{Hidalgo:2016}
N.~Hidalgo, D.~Wladdimiro, E.~Rosas, Self-adaptive processing graph with
  operator fission for elastic stream processing, Journal of Systems and
  SoftwareIn Press.

\bibitem{Pahl:2015}
C.~Pahl, B.~Lee, Containers and clusters for edge cloud architectures -- a
  technology review, in: 3rd International Conference on Future Internet of
  Things and Cloud, 2015, pp. 379--386.

\bibitem{Ottenwalder:2013}
B.~Ottenw\"{a}lder, B.~Koldehofe, K.~Rothermel, U.~Ramachandran, {MigCEP}:
  Operator migration for mobility driven distributed complex event processing,
  in: 7th {ACM} International Conference on Distributed Event-based Systems,
  DEBS '13, ACM, New York, USA, 2013, pp. 183--194.

\bibitem{Dabek:2004}
F.~Dabek, R.~Cox, F.~Kaashoek, R.~Morris, Vivaldi: A decentralized network
  coordinate system, in: Conference on Applications, Technologies,
  Architectures, and Protocols for Computer Communications, SIGCOMM '04, ACM,
  New York, USA, 2004, pp. 15--26.

\bibitem{ZhuBurstDetection:2003}
Y.~Zhu, D.~Shasha, Efficient elastic burst detection in data streams, in:
  Proceedings of the Ninth ACM SIGKDD International Conference on Knowledge
  Discovery and Data Mining, KDD '03, ACM, New York, USA, 2003, pp. 336--345.

\bibitem{TranChangeDetection:2014}
D.-H. Tran, M.~M. Gaber, K.-U. Sattler, Change detection in streaming data in
  the era of big data: Models and issues, {SIGKDD} Explor. Newsl. 16~(1) (2014)
  30--38.

\bibitem{SarkarAssessmentFog:2015}
S.~Sarkar, S.~Chatterjee, S.~Misra, Assessment of the suitability of fog
  computing in the context of internet of things, {IEEE} Transactions on Cloud
  Computing PP~(99) (2015) 1--1.

\bibitem{SatyanarayananEdge:2017}
M.~Satyanarayanan, Edge computing: Vision and challenges, {USENIX} Association,
  Santa Clara, USA, 2017.

\bibitem{Cardellini:2015}
V.~Cardellini, V.~Grassi, F.~L. Presti, M.~Nardelli, Distributed {QoS}-aware
  scheduling in {Storm}, in: 9th ACM International Conference on Distributed
  Event-Based Systems, DEBS '15, ACM, New York, USA, 2015, pp. 344--347.

\bibitem{TudoranJetStream:2016}
R.~Tudoran, A.~Costan, O.~Nano, I.~Santos, H.~Soncu, G.~Antoniu, Jetstream:
  Enabling high throughput live event streaming on multi-site clouds, Future
  Generation Computer Systems 54 (2016) 274--291.
\newblock \href
  {http://dx.doi.org/http://dx.doi.org/10.1016/j.future.2015.01.016}
  {\path{doi:http://dx.doi.org/10.1016/j.future.2015.01.016}}.

\bibitem{MoralesFogStream:2014}
J.~Morales, E.~Rosas, N.~Hidalgo, Symbiosis: Sharing mobile resources for
  stream processing, in: {IEEE} Symposium on Computers and Communications
  ({ISCC 2014}), Vol. Workshops, 2014, pp. 1--6.

\bibitem{Pahl:2016}
C.~Pahl, S.~Helmer, L.~Miori, J.~Sanin, B.~Lee, A container-based edge cloud
  paas architecture based on raspberry pi clusters, in: {IEEE} 4th Int. Conf.
  on Future Internet of Things and Cloud Workshops ({FiCloudW}), 2016, pp.
  117--124.

\bibitem{Ismail:2015}
B.~I. Ismail, E.~M. Goortani, M.~B.~A. Karim, W.~M. Tat, S.~Setapa, J.~Y. Luke,
  O.~H. Hoe, Evaluation of docker as edge computing platform, in: IEEE
  Conference on Open Systems ({ICOS 2015}), 2015, pp. 130--135.

\bibitem{Yangui:2016}
S.~Yangui, P.~Ravindran, O.~Bibani, R.~H. Glitho, N.~B. Hadj-Alouane, M.~J.
  Morrow, P.~A. Polakos, A platform as-a-service for hybrid cloud/fog
  environments, in: {IEEE} International Symposium on Local and Metropolitan
  Area Networks (LANMAN), 2016, pp. 1--7.

\bibitem{CloudFoundry}
{Cloud Foundry}, https://www.cloudfoundry.org/ (2016).

\bibitem{Morabito:2016}
R.~Morabito, N.~Beijar, Enabling data processing at the network edge through
  lightweight virtualization technologies, in: 2016 {IEEE} International
  Conference on Sensing, Communication and Networking (SECON Workshops), 2016,
  pp. 1--6.

\bibitem{Novo:2015}
O.~Novo, N.~Beijar, M.~Ocak, J.~Kjallman, M.~Komu, T.~Kauppinen, Capillary
  networks - bridging the cellular and iot worlds, in: {IEEE} 2nd World Forum
  on Internet of Things ({WF-IoT}), 2015, pp. 571--578.

\bibitem{Petrolo:2016}
R.~Petrolo, R.~Morabito, V.~Loscr{\`i}, N.~Mitton, The design of the gateway
  for the cloud of things, Annals of Telecommunications (2016) 1--10.

\bibitem{Hochreiner:2016}
C.~Hochreiner, M.~Vogler, P.~Waibel, S.~Dustdar, {VISP}: An ecosystem for
  elastic data stream processing for the internet of things, in: 20th {IEEE}
  International Enterprise Distributed Object Computing Conference ({EDOC
  2016}), 2016, pp. 1--11.

\bibitem{GaiCloudlet:2016}
K.~Gai, M.~Qiu, H.~Zhao, L.~Tao, Z.~Zong,
  \href{http://www.sciencedirect.com/science/article/pii/S108480451500123X}{Dynamic
  energy-aware cloudlet-based mobile cloud computing model for green
  computing}, Journal of Network and Computer Applications 59~(Supplement C)
  (2016) 46--54.
\newblock \href {http://dx.doi.org/https://doi.org/10.1016/j.jnca.2015.05.016}
  {\path{doi:https://doi.org/10.1016/j.jnca.2015.05.016}}.
\newline\urlprefix\url{http://www.sciencedirect.com/science/article/pii/S108480451500123X}

\bibitem{GaiMemory:2016}
K.~Gai, M.~Qiu, H.~Zhao, Cost-aware multimedia data allocation for
  heterogeneous memory using genetic algorithm in cloud computing, IEEE
  Transactions on Cloud Computing PP~(99) (2016) 1--1.
\newblock \href {http://dx.doi.org/10.1109/TCC.2016.2594172}
  {\path{doi:10.1109/TCC.2016.2594172}}.

\bibitem{BenoitLinearChain:2013}
A.~Benoit, A.~Dobrila, J.-M. Nicod, L.~Philippe, Scheduling linear chain
  streaming applications on heterogeneous systems with failures, Future
  Generation Computer Systems 29~(5) (2013) 1140--1151, special section: Hybrid
  Cloud Computing.
\newblock \href
  {http://dx.doi.org/http://dx.doi.org/10.1016/j.future.2012.12.015}
  {\path{doi:http://dx.doi.org/10.1016/j.future.2012.12.015}}.

\bibitem{RohPlacement:2017}
H.~Roh, C.~Jung, K.~Kim, S.~Pack, W.~Lee,
  \href{http://www.sciencedirect.com/science/article/pii/S1084804516303101}{Joint
  flow and virtual machine placement in hybrid cloud data centers}, Journal of
  Network and Computer Applications 85 (2017) 4--13, intelligent Systems for
  Heterogeneous Networks.
\newblock \href {http://dx.doi.org/https://doi.org/10.1016/j.jnca.2016.12.006}
  {\path{doi:https://doi.org/10.1016/j.jnca.2016.12.006}}.
\newline\urlprefix\url{http://www.sciencedirect.com/science/article/pii/S1084804516303101}

\bibitem{Cardellini:2016}
V.~Cardellini, V.~Grassi, F.~L. Presti, M.~Nardelli, Optimal operator placement
  for distributed stream processing applications, in: 10th {ACM} International
  Conference on Distributed and Event-based Systems, DEBS '16, ACM, New York,
  USA, 2016, pp. 69--80.

\bibitem{GuCostStream:2016}
L.~Gu, D.~Zeng, S.~Guo, Y.~Xiang, J.~Hu, A general communication cost
  optimization framework for big data stream processing in geo-distributed data
  centers, {IEEE} Transactions on Computers 65~(1) (2016) 19--29.

\bibitem{Tziritas:2016}
N.~Tziritas, T.~Loukopoulos, S.~U. Khan, C.~Z. Xu, A.~Y. Zomaya, On improving
  constrained single and group operator placement using evictions in big data
  environments, {IEEE} Transactions on Services Computing 9~(5) (2016)
  818--831.

\bibitem{ChenStreamCost:2016}
W.~Chen, I.~Paik, Z.~Li, Cost-aware streaming workflow allocation on
  geo-distributed data centers, {IEEE} Transactions on ComputersTo appear.

\bibitem{Mehdipour:2016}
F.~Mehdipour, B.~Javadi, A.~Mahanti, {FOG-Engine}: Towards big data analytics
  in the fog, in: {IEEE} 14th Intl Conf on Dependable, Autonomic and Secure
  Computing, 14th Int. Conf on Pervasive Intelligence and Computing, 2nd Int.
  Conf on Big Data Intelligence and Computing and Cyber Science and Technology
  Congress (DASC/PiCom/DataCom/CyberSciTech), 2016, pp. 640--646.

\bibitem{Shen:2015}
Z.~Shen, V.~Kumaran, M.~J. Franklin, S.~Krishnamurthy, A.~Bhat, M.~Kumar,
  R.~Lerche, K.~Macpherson, {CSA}: Streaming engine for internet of things,
  IEEE Data Eng. Bull. 38~(4) (2015) 39--50.

\bibitem{ChengEdgeStream:2016}
B.~Cheng, A.~Papageorgiou, M.~Bauer, Geelytics: Enabling on-demand edge
  analytics over scoped data sources, in: {IEEE} International Congress on Big
  Data ({BigData Congress}), 2016, pp. 101--108.

\bibitem{KreutzSDNSurvey:2015}
D.~Kreutz, F.~M.~V. Ramos, P.~E. Verissimo, C.~E. Rothenberg, S.~Azodolmolky,
  S.~Uhlig, Software-defined networking: A comprehensive survey, Proceedings of
  the {IEEE} 103~(1) (2015) 14--76.

\bibitem{VulimiriWANalyticsNSDI:2015}
A.~Vulimiri, C.~Curino, P.~B. Godfrey, T.~Jungblut, J.~Padhye, G.~Varghese,
  Global analytics in the face of bandwidth and regulatory constraints, in:
  12th {USENIX} Symposium on Networked Systems Design and Implementation ({NSDI
  15}), USENIX Association, Oakland, USA, 2015, pp. 323--336.

\bibitem{Kubernets}
{Kubernetes}: Production-grade container orchestration, http://kubernetes.io/
  (2015).

\end{thebibliography}
